\documentclass[apjl]{emulateapj}
\usepackage{amsfonts,amsmath,graphicx,natbib,apjfonts,lscape}

\def\nod{\nodata}

\def\cfa{1}
\def\got{2}

\begin{document}

\title{The Radio Activity-Rotation Relation of Ultracool Dwarfs}

\author{
M.~McLean\altaffilmark{\cfa}, 
E.~Berger\altaffilmark{\cfa},
and A.~Reiners\altaffilmark{\got}
}

\altaffiltext{\cfa}{Harvard-Smithsonian Center for Astrophysics, 60
Garden Street, Cambridge, MA 02138}

\altaffiltext{\got}{Universit\"at G\"ottingen, Institut f\"ur
Astrophysik, Friedrich-Hund-Platz 1, 37077 G\"ottingen, Germany}

\begin{abstract} We present a new radio survey of about 100 late-M and
L dwarfs undertaken with the Very Large Array (VLA).  The sample was
chosen to explore the role of rotation in the radio activity of
ultracool dwarfs.  As part of the survey we discovered radio emission
from three new objects: 2MASS J\,$0518113\!-\!310153$ (M6.5), 2MASS
J\,$0952219\!-\!192431$ (M7), and 2MASS J\,$1314203\!+\!132001$ (M7),
and made an additional detection of LP\,349-25 (M8).  Combining the
new sample with results from our previous studies and from the
literature, we compile the largest sample to date of ultracool dwarfs
with radio observations and measured rotation velocities ($167$
objects).  In the spectral type range M0--M6 we find a radio
activity-rotation relation, with saturation at $L_{\rm rad}/L_{\rm
bol}\approx 10^{-7.5}$ above $v{\rm sin}i\approx 5$ km s$^{-1}$,
similar to the relation in H$\alpha$ and X-rays.  However, at spectral
types $\gtrsim {\rm M7}$ the ratio of radio to bolometric luminosity
increases significantly regardless of rotation velocity, and the
scatter in radio luminosity increases.  In particular, while the most
rapid rotators ($v{\rm sin}i\gtrsim 20$ km s$^{-1}$) exhibit
``super-saturation'' in X-rays and H$\alpha$, this effect is not seen
in the radio.  We also find that ultracool dwarfs with $v{\rm sin}i
\gtrsim 20$ km s$^{-1}$ have a higher radio detection fraction by
about a factor of 3 compared to objects with $v{\rm sin}i\lesssim 10$
km s$^{-1}$.  When measured in terms of the Rossby number ($Ro$), the
radio activity-rotation relation follows a single trend and with no
apparent saturation from G to L dwarfs and down to $Ro\sim 10^{-3}$;
in X-rays and H$\alpha$ there is clear saturation at $Ro\lesssim 0.1$,
with super-saturation beyond M7.  A similar trend is observed for the
radio surface flux ($L_{\rm rad}/R_*^2$) as a function of $Ro$.  The
continued role of rotation in the overall level of radio activity and
in the fraction of active sources, and the single trend of $L_{\rm
rad}/L_{\rm bol}$ and $L_{\rm rad}/R_*^2$ as a function of $Ro$ from G
to L dwarfs indicates that rotation effects are important in
regulating the topology or strength of magnetic fields in at least
some fully-convective dwarfs.  The fact that not all rapid rotators
are detected in the radio provides additional support to the idea of
dual dynamo states proposed from spectropolarimetric observations.
\end{abstract}
 

\section{Introduction}
\label{sec:intro}

Rotation plays a key role in the magnetic dynamos of cool stars.  The
$\alpha\Omega$ dynamo \citep{par55} is the standard mechanism used to
explain magnetic field generation in sun-like stars.  The combination
of winding of magnetic field lines due to differential rotation, and
twisting by convective motions results in the generation of a magnetic
field whose strength is highly dependent on stellar rotation.  Indeed,
this dynamo mechanism is supported by the observed correlation between
rotation and magnetic activity (indicated by H$\alpha$, \ion{Ca}{2}
H\&K, X-rays, and radio), as well as between rotation and inferred
magnetic field strengths (from Zeeman broadening) in F to early-M
dwarfs \citep{nhb+84,sis+88,jjj+00,dfp+98,pmm+03,bbm+10,rbb09,mdp+10}.
The critical parameter appears to be the Rossby number, $Ro=P/\tau_c$,
where $P$ is the rotation period and $\tau_c$ is the convective
turnover timescale; magnetic activity increases with a decreasing
Rossby number (e.g., \citealt{nhb+84}).  However, since the
$\alpha\Omega$ dynamo operates at the transition layer between the
radiative and convective zones in stars where differential rotation is
maximized, a separate mechanism may be required to account for
magnetic fields in fully-convective dwarfs (spectral types $\gtrsim
{\rm M3}$).

Observationally, H$\alpha$ and X-ray activity measurements demonstrate
that the correlation between rotation and activity continues beyond
the expected transition to full convection \citep{dfp+98,mb03}.
However, since the activity in early- to mid-M dwarfs becomes
saturated at a fairly low rotation rate, corresponding to $v\approx 5$
km s$^{-1}$, few objects display the unsaturated correlation.  An
eventual breakdown in the saturated rotation-activity relation is
observed in ultracool dwarfs (spectral type $\gtrsim {\rm M7}$) in
both X-rays and H$\alpha$ (e.g., \citealt{bm95,mb03,bbg+08}), such
that the activity levels decline precipitously in all objects,
independent of rotation (e.g., \citealt{bbf+10}).  A similar breakdown
is seen in the correlation between rotation and magnetic flux $Bf$
\citep{rb10}; here $B$ is the magnetic field and $f$ is the magnetic
filling factor.  Moreover, late-M dwarfs appear to exhibit distinct
regimes of magnetic field topologies and strengths with no obvious
correlation to the stellar rotation \citep{mdp+10}.

There are also hints of super-saturation among the most rapid rotators
(with $v\gtrsim 20$ km s$^{-1}$) where there appears to be a
particularly strong suppression of H$\alpha$ \citep{rb10} and X-ray
\citep{jjj+00,bbg+08} activity at spectral types beyond M7.  It is
unclear whether the super-saturation phenomenon is related to external
effects such as centrifugal coronal stripping
\citep{ju99,jjj+00,bbg+08}, or to the actual generation of the
magnetic field.  Unfortunately, it is not possible to make direct
magnetic flux measurements in rapidly rotating objects because Zeeman
broadening is masked by rotational broadening \citep{mdp+10,rb10}.

Similarly, there is an inherent difficulty in using H$\alpha$ and
X-ray emission to study field generation in ultracool dwarfs since the
decrease in activity may be due to a decoupling between the
increasingly neutral atmosphere and any existing magnetic fields
\citep{mbs+02}, or to a change in the bulk coronal density (e.g.,
\citealt{bbg+08}).  This is potentially manifested in the sharp
breakdown of the tight radio/X-ray activity correlation
\citep{gb93,bg94} at spectral type M7 \citep{ber02,ber06,bbf+10}.
Thus, while X-ray and H$\alpha$ activity plummet in ultracool dwarfs,
the ratio of radio to bolometric luminosity actually increases in the
coolest objects, while the radio luminosity itself remains largely
unchanged, at least to spectral type $\sim {\rm L4}$
\citep{ber02,ber06,bbf+10}.  Thus, radio observations indicate that
ultracool dwarfs are capable of generating stable large-scale
kilo-Gauss magnetic fields.  This result has been confirmed with
Zeeman broadening observations of M dwarfs \citep{rb10}, as well as
with magnetic topology studies using spectropolarimetry (Zeeman
Doppler Imaging; \citealt{dfc+06,mdp+10}).  However, radio activity
remains a unique tool for studying the field topology and dissipation
in rapid rotators.

Theoretical studies have led to several proposed models to explain the
continued presence of magnetic fields in fully convective stars and
brown dwarfs.  \citet{ddr93} proposed a turbulent dynamo that
generates small-scale chaotic magnetic fields, with little dependence
on rotation.  However, this mechanism cannot produce sufficient
large-scale magnetic fields to explain the observed Zeeman broadening,
or the strong radio activity among ultracool dwarfs, particularly
periodic radio emission that appears to require a substantial field
with low multipole order \citep{brp+09,mbi+11}.  \citet{ck06} explored
the $\alpha^2$ dynamo, which predicts large-scale primarily toroidal
fields with a strong dependence on rotation, and saturation at high
rotation rates where the $\alpha^2$ mode becomes super-critical.  This
dynamo model has been proposed to dominate among all fast solid-body
rotators, even the partially radiative early-M dwarfs.  \citet{dsb06}
conducted 3-dimensional hydrodynamic and magnetohydrodynamic (MHD)
simulations of fully convective spheres, and found magnetic fields on
all spatial scales, as well as differential rotation.  They also found
that the fields on the largest scales increase with rotation rate,
reaching saturation only at fast rotation ($Ro\approx 0.01$), and
exhibiting no sign of super-saturation.  The MHD simulations of
\citet{bro08} also find that faster rotation produces higher magnetic
energy densities, as well as magnetic fields on increasingly
large-scales with a dipolar topology.  However, their models do not
include the fastest rotators.

In this paper, we study the relation between radio activity and
rotation in M and L dwarfs as a way to explore and constrain the
magnetic dynamo mechanism of ultracool dwarfs.  A radio
activity-rotation relation has been found in F--K stars
\citep{sis+88,ss89}, as expected from the X-ray activity-rotation
relation and the strong radio/X-ray correlation.  Based on a small
sample of ultracool dwarfs, \citet{bbg+08} found hints that a
connection between radio activity and rotation may persist in late-M
and L dwarfs, despite the breakdown in the H$\alpha$ and X-ray
activity-rotation relations.  Here we present a much larger sample of
objects, taking advantage of new radio observations of 104 M and L
dwarfs, as well as new rotation velocity measurements for
previously-studied objects \citep{rb08,rb10}.  We present the radio
observations in \S\ref{sec:obs}, and new ultracool dwarf detections in
\S\ref{sec:rad}.  In \S\ref{sec:rot} we study the role of rotation in
producing radio activity, and we discuss implications for dynamo
models in \S\ref{sec:disc}.  Our key finding is that the fastest
rotators (highest $v{\rm sin}i$ and lowest Rossby number) have higher
ratios of radio to bolometric luminosity, higher radio surface fluxes,
and a higher radio detection fraction, suggesting that rotation
continues to play a role in the magnetic dynamo mechanism of ultracool
dwarfs.

\section{Observations}
\label{sec:obs}

\begin{figure}
\epsscale{1}
\plotone{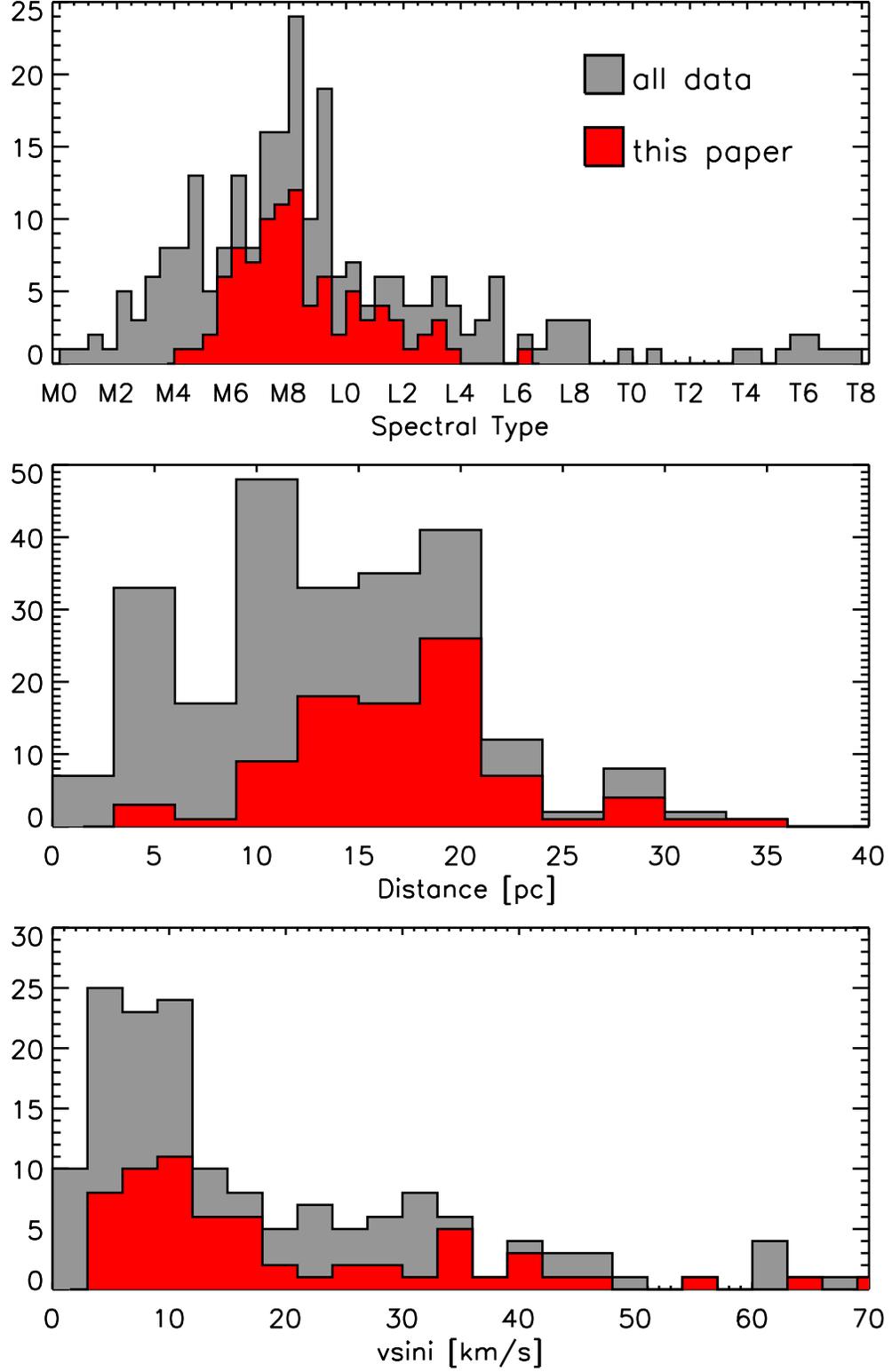}
\caption{Properties of the survey sources, including spectral types
(top), distances (middle), and projected rotation velocities (bottom).
The new objects from this survey (Table~\ref{tab:obsn}) are shown in
red while all data including observations from the literature
(Table~\ref{tab:obso}) are shown in gray.
\label{fig:surv}}
\end{figure}

We carried out a survey of $104$ M and L dwarfs with the Very Large
Array (VLA\footnotemark\footnotetext{The VLA is operated by the
National Radio Astronomy Observatory, a facility of the National
Science Foundation operated under cooperative agreement by Associated
Universities, Inc.}).  The properties of the sources are summarized in
Table~\ref{tab:obsn} and plotted in Figure~\ref{fig:surv}.  The
targets are concentrated in the spectral type range M6--M9, where the
X-ray/radio correlation and the X-ray/H$\alpha$ activity-rotation
relation break down.  The sample includes sources with no previous
radio observations, as well as several objects with previous
detections or upper limits.  The majority of the sample is located at
$\lesssim 20$ pc.  Taking into account objects with new velocity
measurements that were observed in previous radio surveys, we have
increased the number of objects with both radio and rotation
measurements by about a factor of three compared to previous studies.

Each object was observed for 1 hr at 8.46 GHz using the standard
continuum mode with $2\times 50$ MHz contiguous bands.  The flux
density scale was determined using the extragalactic calibrators
3C\,48 (J\,0137+331), 3C\,138 (J\,0521+166), or 3C\,286 (J\,1331+305),
while the phase was monitored using calibrators located within
$10^\circ$ of the target sources.  The data were reduced and analyzed
using the Astronomical Image Processing System.  The resulting flux
density measurements are given in Table~\ref{tab:obsn}.  Previous
radio observations collection from the literature are presented in
Table~\ref{tab:obso}
\citep{wjk89,kll99,bbb+01,ber02,bp05,ber06,pol+07,brp+09}.

\section{New Radio Detections} 
\label{sec:rad}

As part of this new survey we detect radio emission from four objects:
2MASS J\,$0518113-310153$ (M6.5), 2MASS J\,$0952219-192431$ (M7),
2MASS J\,$1314203+1320011$ (M7; \citealt{mbi+11}), and the
previously-detected binary system LP\,349-25 (M8; \citealt{pol+07}).
The measured flux densities are $181\pm 27$ $\mu$Jy, $233\pm 15$
$\mu$Jy, $1156\pm 15$ $\mu$Jy, and $323\pm 14$ $\mu$Jy, respectively.
The fractional circular polarization for the four objects are
$f_c\lesssim 45\%$ (2M\,$0518-3101$), $f_c\lesssim 30\%$
(2M\,$0952-1924$), $f_c=18\pm 2\%$ (2M\,$1314+1320$), and $f_c\lesssim
23\%$ (LP\,349-25).  These values are consistent with the level of
circular polarization in the quiescent emission observed in other
ultracool dwarfs (e.g., \citealt{ber02,ber06}).  The flux density
measured for LP\,349-25 is consistent with the value reported by
\citet{pol+07}.  No radio emission was detected from GJ\,2005,
BRI\,0021-0214 or GJ\,234A, despite previous radio detections
\citep{wjk89,ber02,ber06}.  Our upper limits for GJ\,2005 and
BRI\,0021-0214 are only a factor of 1.7 and 1.4 below the previous
detections, respectively.  However, the upper limit on GJ\,234A is
almost 7 times below the detection from \citet{wjk89}, indicative of
long-term variability.

We also carried out a 10 hr follow-up observation of 2M\,$0952-1924$
at 4.96 GHz and 8.46 GHz, but found no significant detection, to a
limit of $69$ $\mu$Jy, a factor of 2.4 below the original detection.
This indicates that the initial detection was either a flare, or that
the source experiences long-term variability.  Since 2M\,$0952-1924$
has a rotation velocity of $v{\rm sin}i\approx 6$ km s$^{-1}$, its
rotation period could be as long as 20 hr, indicating that the
non-detection in 10 hr could also result from significant rotational
modulation \citep{brr+05,had+06,hbl+07,brp+09,mbi+11}.

\section{Exploring the Role of Rotation}
\label{sec:rot}

\begin{figure}
\epsscale{1}
\plotone{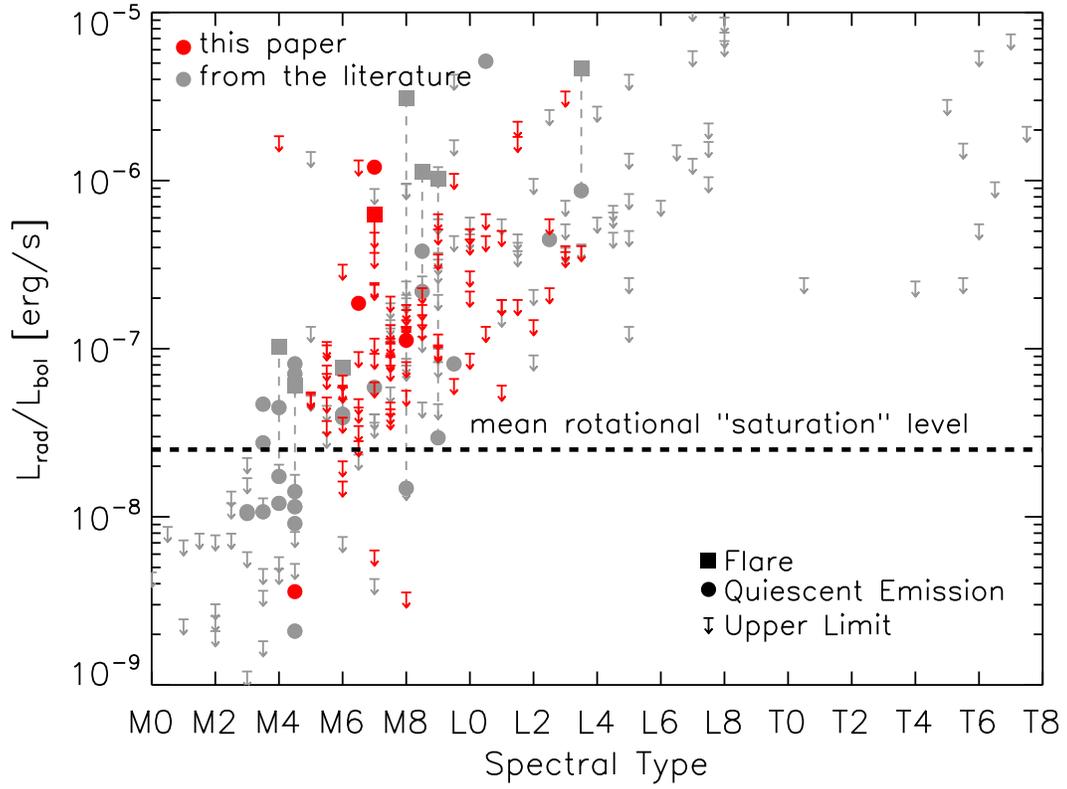}
\caption{Ratio of radio to bolometric luminosity as a function of
spectral type.  Shown are flares (squares), quiescent emission
(circles), and upper limits (arrows).  Red symbols represent the
objects from this survey (Table ~\ref{tab:obsn}) while gray symbols
represent objects from the literature (Table~\ref{tab:obso}).  The
mean rotationally saturated level of emission for early- to mid-M
dwarfs is shown as a dashed line (see Figure~\ref{fig:rbol}).  The
clear trend of increased $L_{\rm rad}/L_{\rm bol}$ as a function of
later spectral type is seen in the ultracool dwarfs.
\label{fig:lrlb}}
\end{figure}

To explore the connection between rotation and radio activity we study
the full sample of M and L dwarfs with radio observations and measured
rotation velocities.  In Figure~\ref{fig:lrlb} we plot the ratio of
radio to bolometric luminosity as a function of spectral type for the
full sample (Tables~\ref{tab:obsn} and \ref{tab:obso}).  We find an
overall trend of increasing radio activity with later spectral type,
at least to spectral type $\sim{\rm L4}$, with a dearth of sources
with $L_{\rm rad}/L_{\rm bol}\gtrsim 10^{-7}$ in spectral types
earlier than M6 (see also \citealt{ber02,ber06}).  Moreover,
essentially every detected object beyond a spectral type of M7
exhibits a value of $L_{\rm rad}/L_{\rm bol}$ that is larger than the
saturated activity level in the M0--M6 dwarfs.

\begin{figure}
\epsscale{1}
\plotone{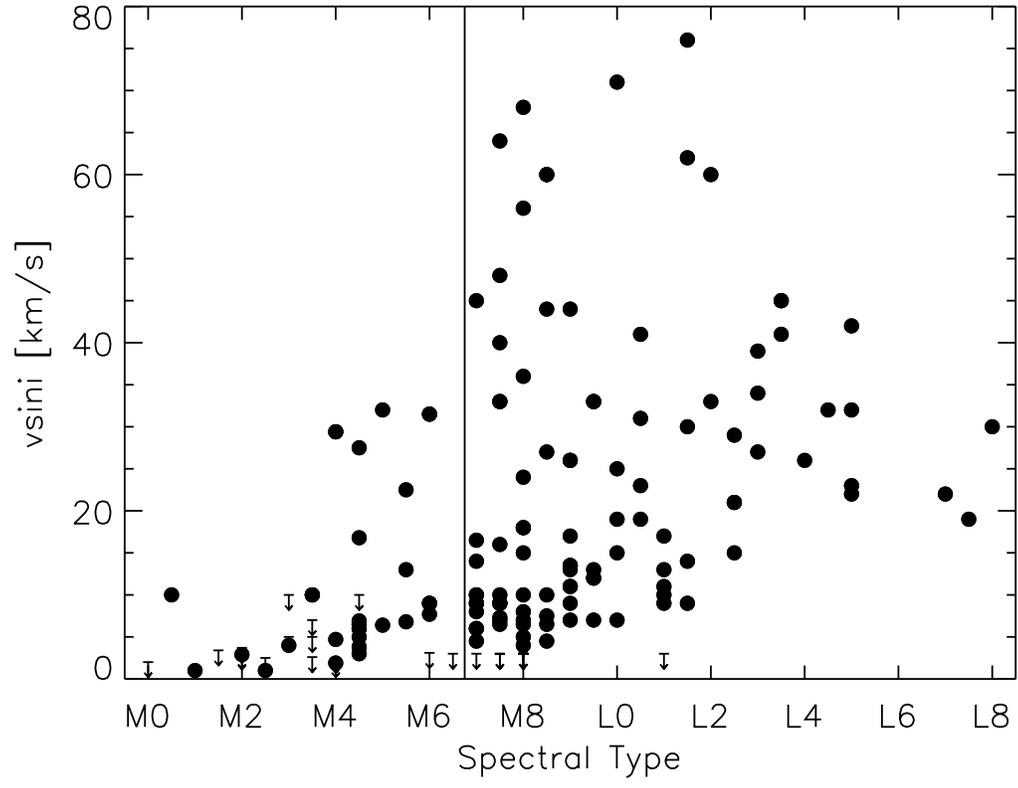}
\caption{Projected rotation velocities ($v\sin i$) as a function of
spectral type for the objects studied in this paper.  The region above
$v{\rm sin}i\approx 30$ km s$^{-1}$ contains no early- or mid-M
dwarfs, but is well populated by objects in the range M7--L5.
\label{fig:vsini}}
\end{figure}

\begin{figure}
\epsscale{1}
\plotone{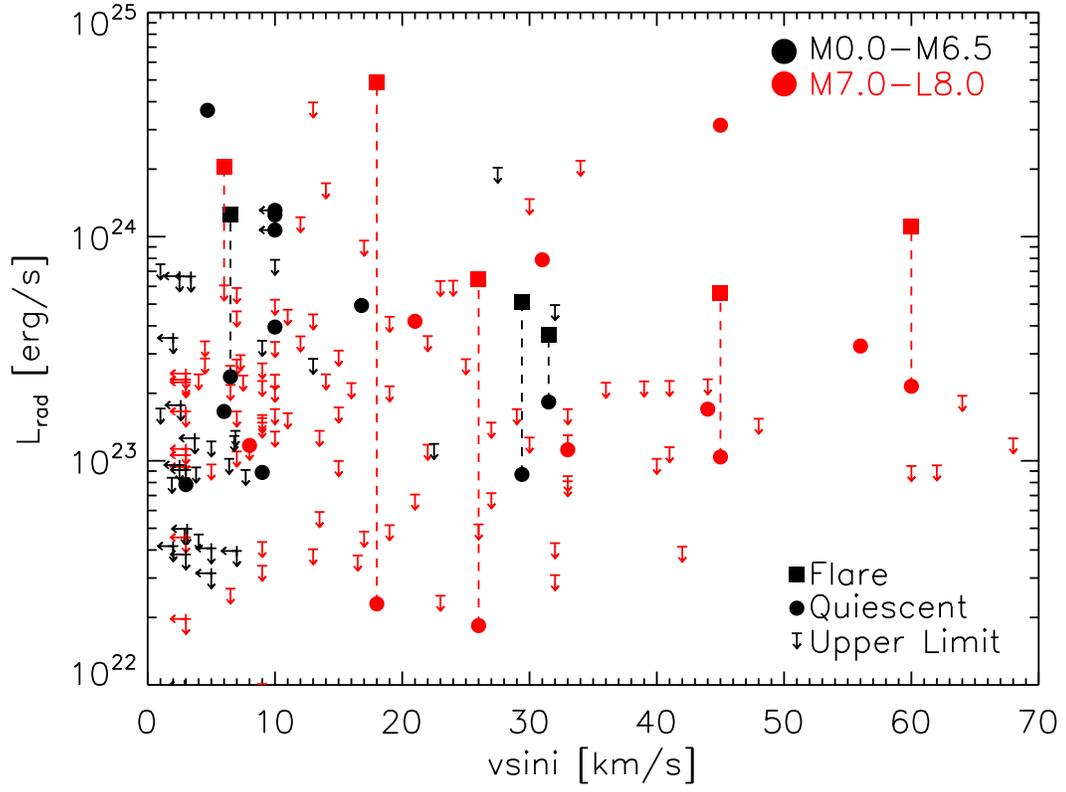}
\caption{Radio luminosity as a function of projected rotation
velocity.  Shown are flares (squares), quiescent emission (circles),
and radio upper limits (arrows).  Left-arrows indicate upper limits in
$v{\rm sin}i$.  Red symbols represent objects later than M7, while
black symbols represent objects with spectral types M0--M6.5.  No
obvious trend is detected, but there is a tantalizing paucity of
objects with radio luminosity of $\lesssim 10^{23}$ erg s$^{-1}$ at
$v{\rm sin}i\gtrsim 30$ km s$^{-1}$.
\label{fig:rrad}}
\end{figure}

The distribution of rotation velocities as a function of spectral type
is shown in Figure~\ref{fig:vsini}.  There are no M0--M6 dwarfs with
rotation velocities of $v{\rm sin}i\gtrsim 30$ km s$^{-1}$, while
among the ultracool dwarfs the sample is fairly uniformly distributed
over the range of $\approx 5-60$ km s$^{-1}$.  Combining the rotation
velocities with the radio luminosities (Figure~\ref{fig:rrad}), we
find no clear correlation, although there is a tantalizing paucity of
objects with $v{\rm sin}i\gtrsim 30$ km s$^{-1}$ and radio luminosity
of $\lesssim 10^{23}$ erg s$^{-1}$, which are present at $v{\rm
sin}i\lesssim 30$ km s$^{-1}$.  The lack of an obvious change in radio
luminosity from early-M dwarfs to ultracool dwarfs contrasts with the
trends seen in H$\alpha$ and X-rays \citep{bbf+10}.

\begin{figure}
\epsscale{1}
\plotone{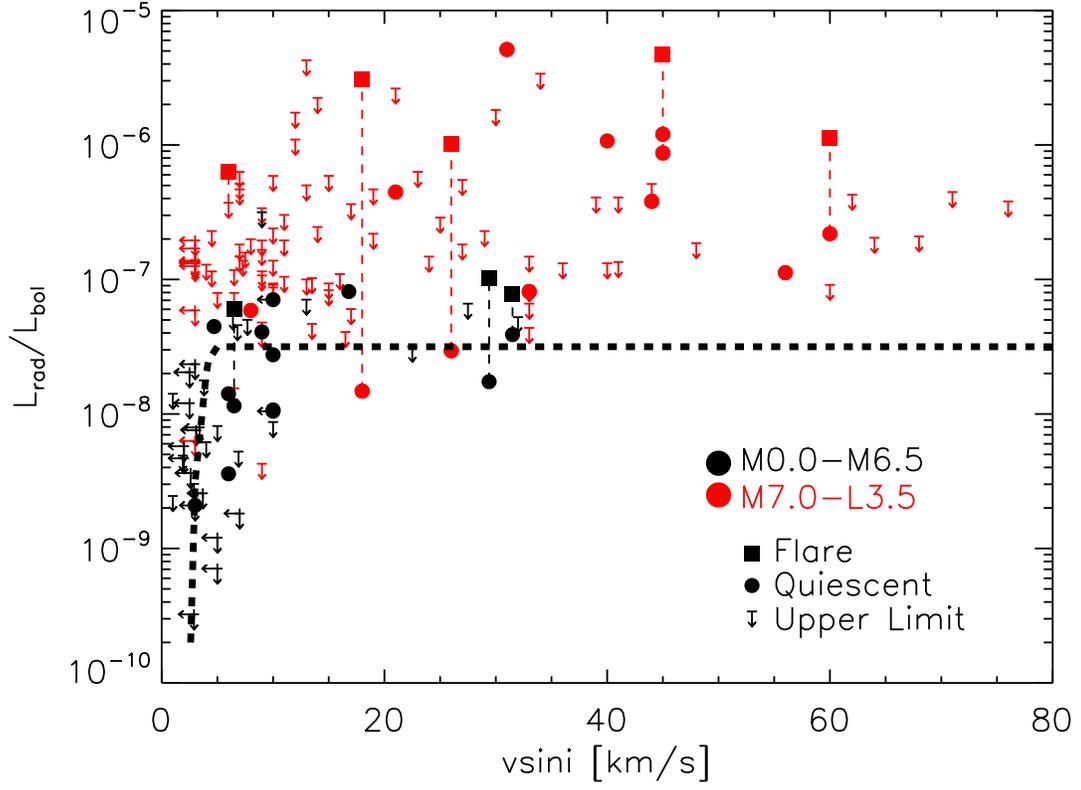}
\caption{Ratio of radio to bolometric luminosity as a function of
projected rotation velocity.  Shown are flares (squares), quiescent
emission (circles), and radio upper limits (arrows).  Left-arrows
indicate upper limits in $v{\rm sin}i$.  Red symbols represent objects
later than M7, while black symbols represent objects with spectral
types M0--M6.5.  The early- to mid-M dwarfs appear to reach a
saturated level of $L_{\rm rad}/L_{\rm bol}\approx 10^{-7.5}$ at
$v{\rm sin}i\gtrsim 5$ km$^{-1}$.  The black dashed line indicates a
rough fit to the radio activity-rotation relation for objects earlier
than M7.  There is a general increase in $L_{\rm rad}/L_{\rm bol}$
among the late-M and L dwarfs (as seen in Figure~\ref{fig:lrlb}), as
well as an increase in the scatter.  There is also an indications of a
trend towards higher luminosity ratios in the fastest rotators, $v{\rm
sin}i\gtrsim 30$ km s$^{-1}$.
\label{fig:rbol}}
\end{figure}

\begin{figure}
\epsscale{1}
\plotone{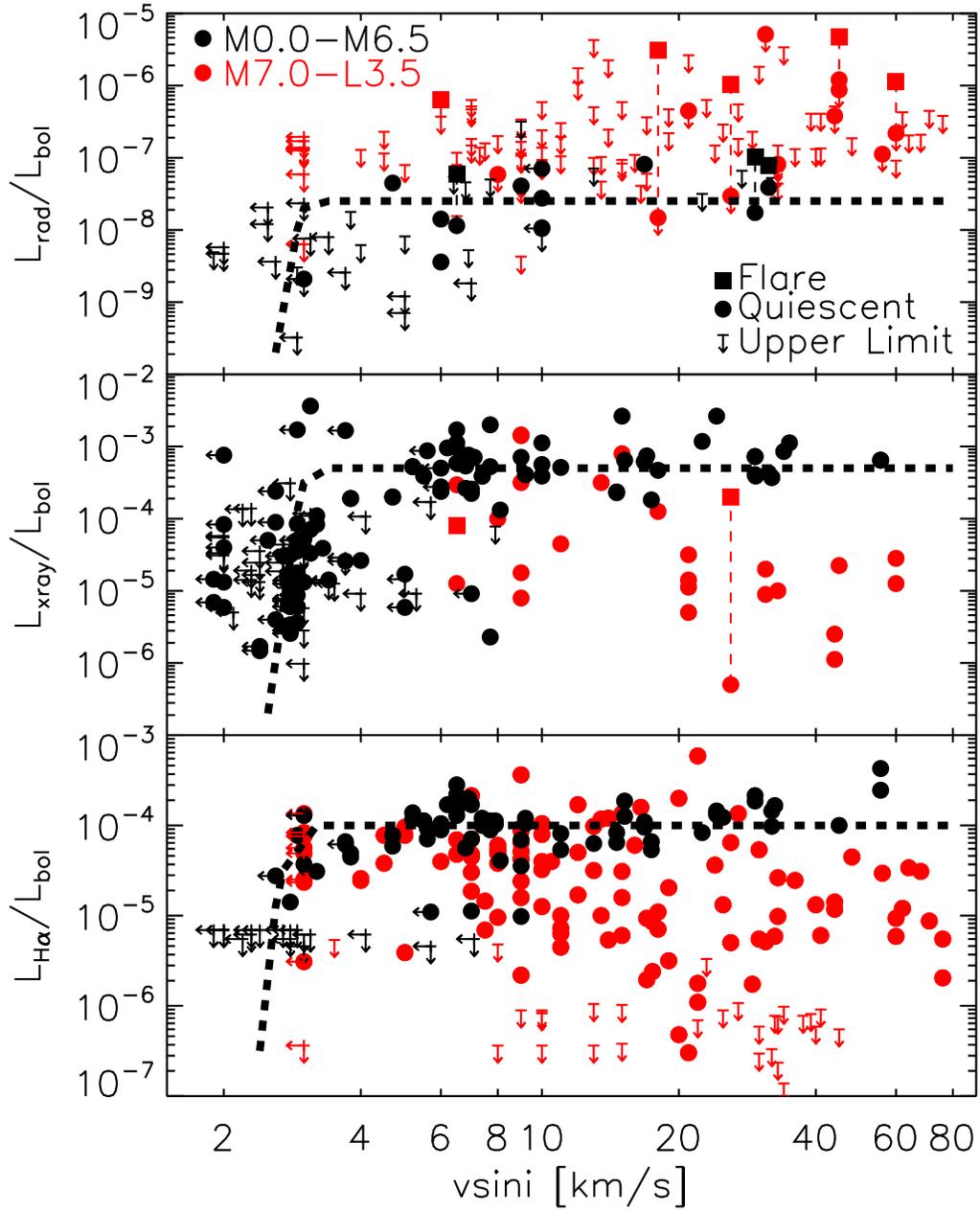}
\caption{Radio, X-ray and H$\alpha$ activity as a function of
projected rotation velocity.  Shown are flares (squares), quiescent
emission (circles), and upper limits (arrows). Left-arrows indicate
upper limits in $v{\rm sin}i$.  Red symbols represent objects later
than M7, while black symbols represent objects with spectral types
M0--M6.5.  X-ray data are taken from \citet{bbg+08}, \citet{jjj+00},
\citet{dfp+98}, and \citet{pmm+03}.  H$\alpha$ data are taken from
\citet{dfp+98}, \citet{mb03}, \citet{rb08}, and \citet{rb10}.  All
three activity indicators appear to saturate at $\sim 5$ km s$^{-1}$
for objects earlier than M6.  The H$\alpha$ and X-ray
activity-rotation relations break down at M7--M9 \citep{bbg+08,rb10}
with the activity dropping rapidly and the scatter increasing in later
spectral types.  There are hints of ``super-saturation'' above $\sim
30$ km s$^{-1}$ where the luminosity drops in the most rapidly
rotating objects.  The radio activity follows a similar trend to the
X-ray activity in the earlier objects.  However, beyond M7, it begins
to exhibit the opposite behavior.
\label{fig:rmulti}}
\end{figure}

Since the X-ray and H$\alpha$ rotation trends are strongest when
scaled relative to the bolometric luminosity, we plot $L_{\rm rad}/
L_{\rm bol}$ as a function of rotation velocity in
Figure~\ref{fig:rbol}.  In the early- to mid-M dwarfs we find an
apparent radio rotation-activity relation, with subsequent saturation
at $v{\rm sin}i\gtrsim 5$ km s$^{-1}$ and $L_{\rm rad}/L_{\rm bol}
\approx 10^{-7.5}$.  There are few detections below the saturation
velocity, but the bulk of the upper limits for the slow rotators are
well below the saturated emission level.  This behavior is consistent
with the rotation-activity relation observed in the X-rays, as
expected from the radio/X-ray correlation in early-M dwarfs
\citep{gb93,bg94}.  It is also similar to the H$\alpha$
rotation-activity relation \citep{dfp+98,mb03}.  On the other hand,
the detected late-M and L dwarfs exhibit a general increase in $L_{\rm
rad}/L_{\rm bol}$ compared to M0--M6 (Figure~\ref{fig:lrlb}).
Therefore, the ultracool dwarfs no longer follow the saturation level
observed in the early-M dwarfs, and instead reside at higher values of
$L_{\rm rad}/L_{\rm bol}\sim 10^{-6.4}$.  There is also an increase in
the scatter of radio activity levels in ultracool dwarfs, similar to
that seen in X-rays and H$\alpha$ (see Figure~\ref{fig:rmulti}).  The
increased scatter is indicative of a breakdown in the correlation
between the activity level and rotation velocity.  On the other hand,
at $v{\rm sin}i\gtrsim 20$ km s$^{-1}$, where the X-ray and H$\alpha$
activity appear to exhibit super-saturation, there are indications of
a trend towards higher radio activity levels
(Figure~\ref{fig:rmulti}).  It is therefore clear that the radio
activity and the H$\alpha$/X-ray trends diverge in ultracool dwarfs,
regardless of whether we normalize by the bolometric luminosity or
not.

\begin{figure}
\epsscale{1}
\plotone{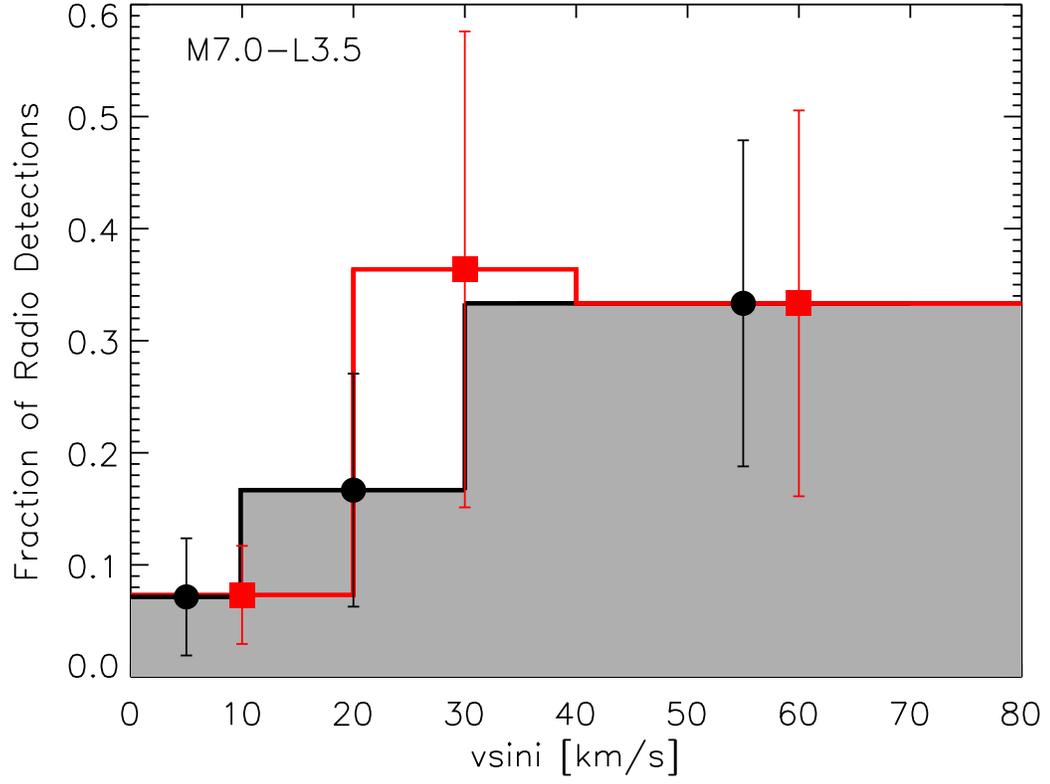}
\caption{Fraction of objects with radio detections as a function of
rotation velocity for ultracool dwarfs with spectral types M7--L4.
Upper limits above $L_{\rm rad}\approx 2.5\times 10^{23}$ erg s$^{-1}$
have been excluded.  Uncertainties are determined from the Poisson
distribution.  Two different sets of binning are shown in order to
test the impact of the choice of boundaries.  There is a clear
increase in the fraction of detected objects among the fastest
rotators.  The transition seems to occur at $v{\rm sin}i\approx 20-30$
km s$^{-1}$.
\label{fig:rstat}}
\end{figure}

We further explore the role of rotation by investigating the fraction
of objects with radio detections as a function of rotation velocity.
We divide the objects with spectral types M7--L4 into three $v{\rm
sin}i$ bins, using two sets of binning, and retaining only significant
non-detections, i.e., those with $L_{\rm rad}\lesssim 2.5\times
10^{23}$ erg s$^{-1}$, which is the typical luminosity of the detected
sources.  The results are shown in Figure~\ref{fig:rstat}.  For both
sets of binning we find a clear increase in the fraction of radio
detections as a function of $v{\rm sin}i$, from a few percent at
$v{\rm sin}i\lesssim 15$ km s$^{-1}$ to about $30\%$ at $v{\rm
sin}i\gtrsim 30$ km s$^{-1}$.  This result suggests that while radio
luminosity may not increase with faster rotation, the probability of
producing radio emission does depend on fast rotation.  This may be
due to the influence of rotation on the magnetic field strength and/or
its topology.

\begin{figure}
\epsscale{1}
\plotone{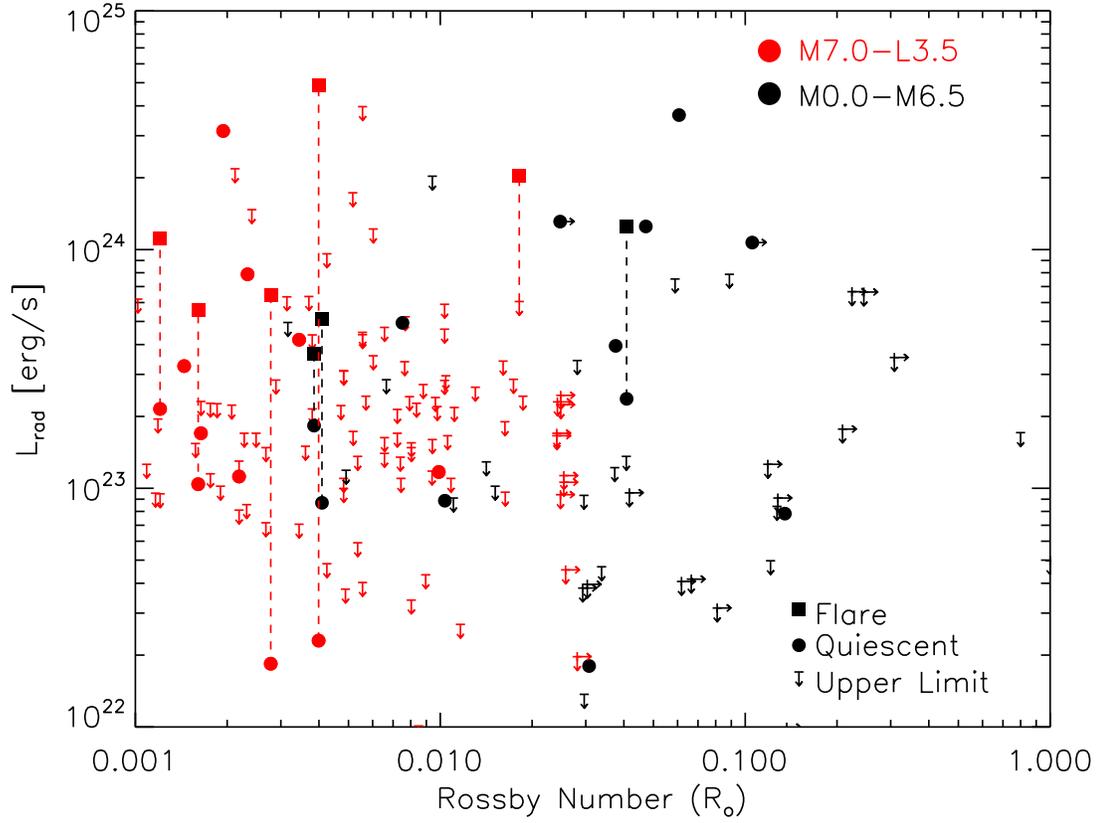}
\caption{Radio luminosity as a function of Rossby number
($Ro=P/\tau_c$).  Shown are flares (squares), quiescent emission
(circles), and radio upper limits (arrows).  Right-arrows indicate
lower limits in $Ro$.  Red symbols represent objects later than M7,
while black symbols represent objects with spectral types M0--M6.5.
As in Figure~\ref{fig:rrad}, no correlation between $L_{\rm rad}$ and
$Ro$ is obvious, but the bulk of ultracool dwarfs with radio
detections are concentrated at low Rossby numbers, $Ro\lesssim 5\times
10^{-3}$.
\label{fig:roslum}}
\end{figure}

Studies of X-ray, \ion{Ca}{2} H\&K, and H$\alpha$ activity indicate
that the Rossby number is the rotation parameter most highly closely
correlated with magnetic activity \citep{nhb+84}.  We estimate the
Rossby numbers for our sample using the method of \citet{rb10}.  The
periods are estimated from $v{\rm sin}i$ combined with radii estimated
from the mass-magnitude \citep{dfs+00} and mass-radius \citep{bca+98}
relations.  A radius of 0.1 $R_\odot$ is used for spectral types
beyond M8.  We estimate $\tau_c$ using the empirical relation of
\citet{ks07} imposing a maximum of 70 d, consistent with
\citet{gil86}.  In Figure~\ref{fig:roslum} we plot radio luminosity as
a function of $Ro$.  As in the case of radio luminosity versus $v{\rm
sin}i$, we find no significant evolution from early-M to ultracool
dwarfs.  However, we note that the ultracool dwarfs with radio
emission are clearly concentrated at $Ro\lesssim 5\times 10^{-3}$,
indicating that objects with low Rossby numbers are more likely to
produce detectable radio activity.

\begin{figure}
\epsscale{1}
\plotone{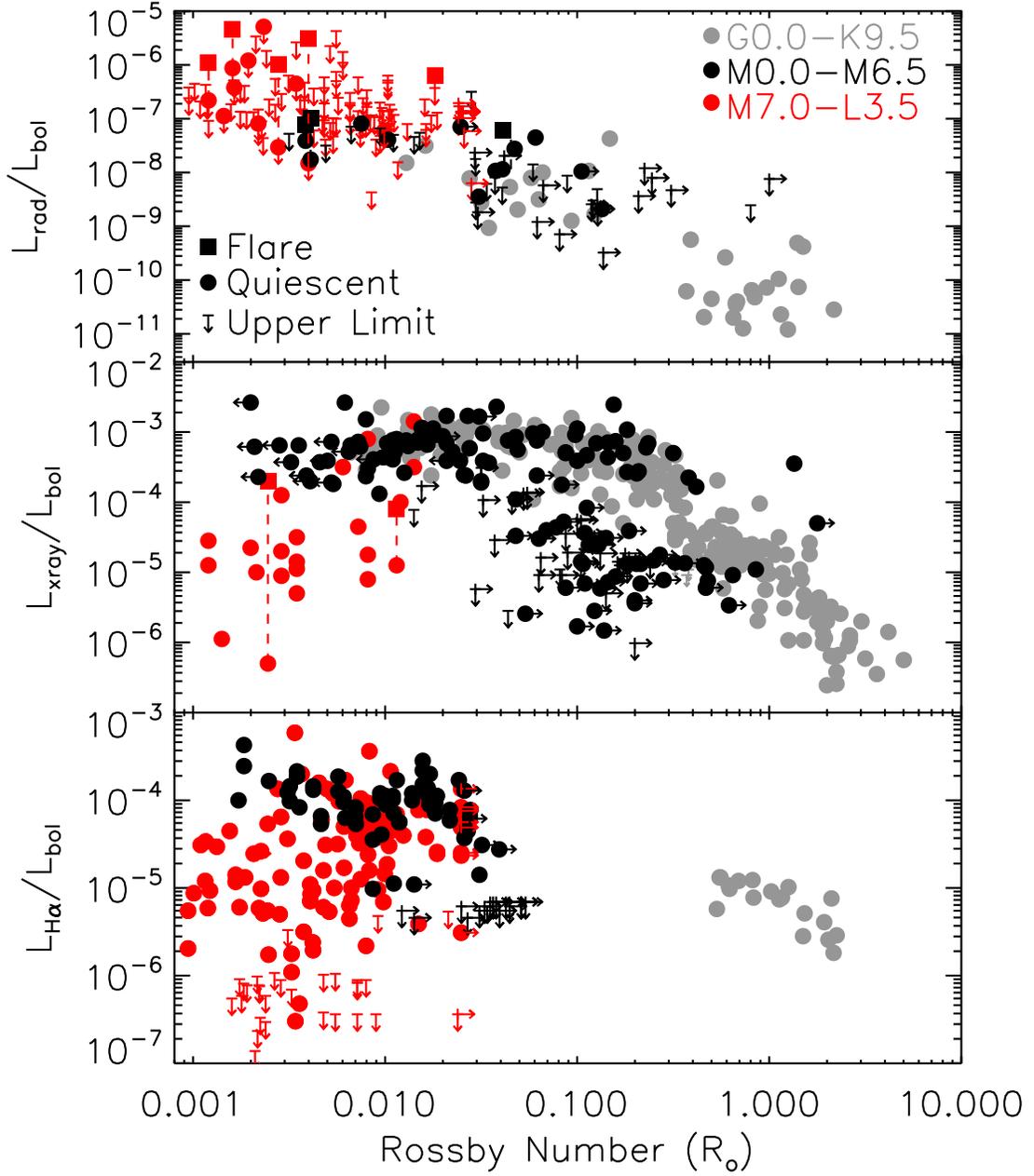}
\caption{Radio, X-ray and H$\alpha$ activity as a function of Rossby
number.  Shown are flares (squares), quiescent emission (circles), and
upper limits (arrows). Right-arrows indicate lower limits in $Ro$.
Red symbols represent objects later than M7, black symbols represent
objects with spectral types M0--M6.5, and gray symbols represent
spectral types G--K.  X-ray data are from \citet{jjj+00},
\citet{dfp+98}, \citet{pmm+03}, and \citet{bbg+08}.  H$\alpha$ data
are from \citet{mek85}, \citet{dfp+98}, \citet{mb03}, \citet{rb08},
and \citet{rb10}.  Radio data for the G--K stars are from
\citet{sis+88}, \citet{ss89}, and \citet{gsb95}.  As in Figure
~\ref{fig:rmulti}, the H$\alpha$ and X-ray activity-rotation relations
break down at M7--M9 \citep{bbg+08,rb10}, with the activity dropping
rapidly and the scatter increasing in later spectral types.  There are
hints of ``super-saturation'' in the most rapidly rotating ultracool
dwarfs with $Ro\lesssim 0.01$.  The radio activity appears to follow a
single trend with Rossby number from M0--L4 and $Ro\approx
0.1-10^{-3}$.
\label{fig:rosmulti}} 
\end{figure}

In Figure~\ref{fig:rosmulti} we plot the luminosity in radio, X-ray
and H$\alpha$ scaled by the bolometric luminosity as a function of
$Ro$.  We supplement our data with results for F--K stars from the
literature \citep{mek85,sis+88,ss89,dfp+98,jjj+00,pmm+03}.  The
previously-discussed trends in X-ray and H$\alpha$ activity versus
$v{\rm sin}i$ are more pronounced when plotted versus Rossby number.
In particular, for spectral types earlier than M6 the X-ray activity
exhibits a rapid increase by about 3 orders of magnitude as $Ro$
decreases from $\sim 2$ to $\sim 0.2$, followed by saturation at
$Ro\lesssim 0.2$.  The ultracool dwarfs exhibit a clear
super-saturation trend of decreasing activity as a function of
decreasing Rossby number in the range $Ro\approx 10^{-2}-10^{-3}$ (see
also \citealt{bbg+08}).  A similar trend is apparent in H$\alpha$
activity (see also \citealt{rb10}).  On the other hand, in the radio
band we find a uniform trend of increasing activity as a function of
decreasing Rossby number over the range $Ro\approx 0.1-10^{-3}$ and
for spectral types G to L, indicating that at least in some ultracool
dwarfs, there is no evidence for a breakdown in the activity-Rossby
number relation.  A Spearman's rank correlation test for the detected
sources gives $\rho\approx -0.88$ with a null hypothesis (no
correlation) probability of only $\approx 1.1\times 10^{-15}$.  A
linear regression fit indicates an overall trend of $L_{\rm rad}/
L_{\rm bol}\propto Ro^{-1.1}$.

\begin{figure}
\epsscale{1}
\plotone{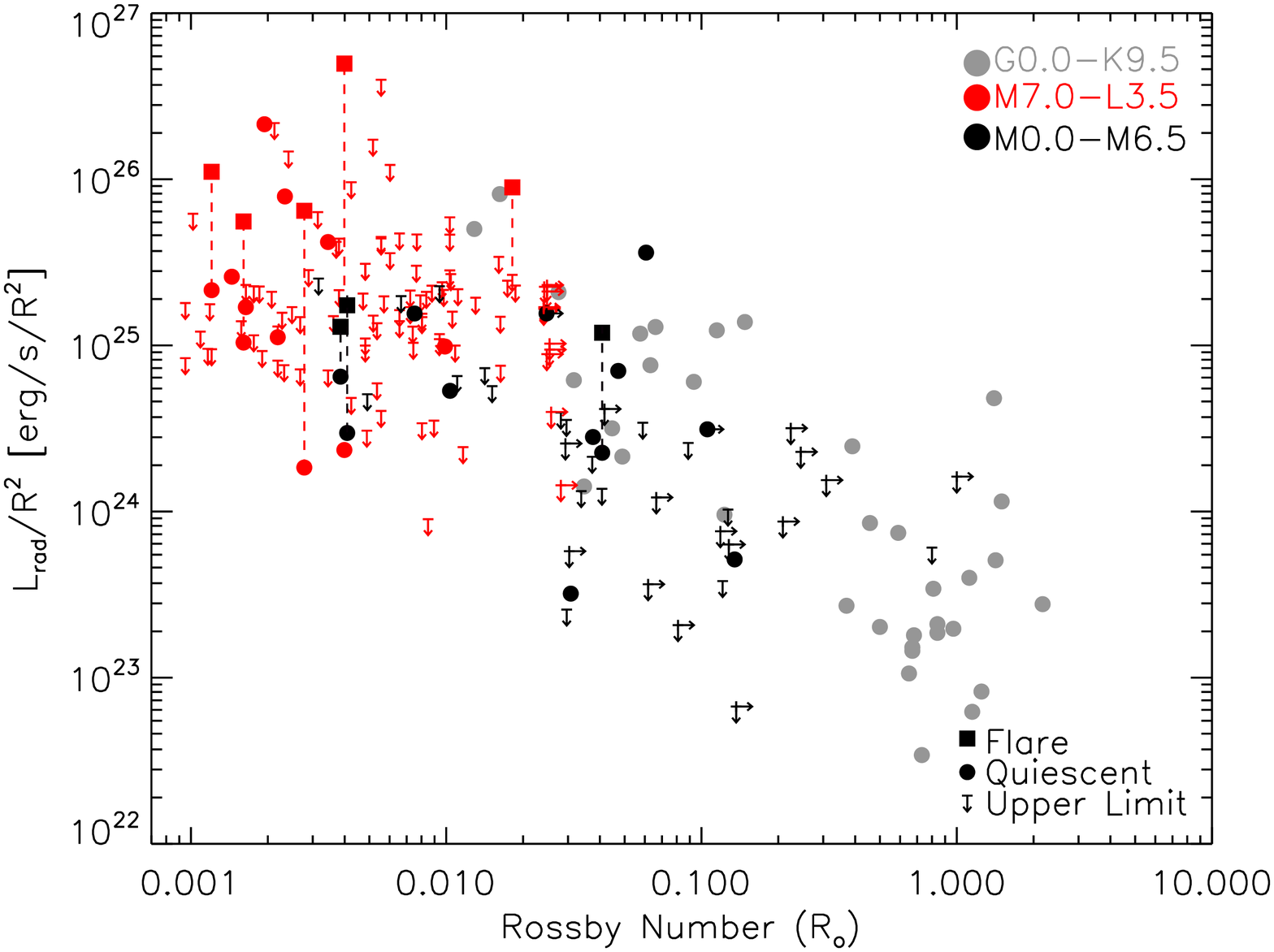}
\caption{Radio surface flux as a function of Rossby number.  Shown are
flares (squares), quiescent emission (circles), and upper limits
(arrows).  Right-arrows indicate lower limits in $Ro$.  Red symbols
represent objects later than M7, black symbols represent objects with
spectral types M0--M6.5, and gray symbols represent spectral types
G--K \citep{sis+88,ss89,gsb95}.  The stellar radii are in units of
$R_\odot$.  As in the case of $L_{\rm rad}/L_{\rm bol}$, the surface
flux appears to follow a single trend from spectral types G2 to L4.
The overall trend is roughly linear, $L_{\rm rad}/R_*^2\propto
Ro^{-1}$.
\label{fig:rosflux}} 
\end{figure}

Finally, \citet{ss89} noted the potential importance of radio surface
flux ($L_{\rm rad}/R_*^2$) as a quantity strongly correlated with
rotation in G--K stars; here $R_*$ is the stellar radius normalized to
solar units.  These authors found that $L_{\rm rad}/R_*^2 \propto
P^{-1.8\pm 0.3}\,R_*^{1.8\pm 0.4}$.  In Figure~\ref{fig:rosflux} we
plot radio surface flux as a function of Rossby number for the
ultracool dwarfs in our sample and from the literature, as well as for
the main sequence stars in the \citet{ss89} sample.  The objects,
ranging from spectral type G to L, again appear to follow a single
trend with respect to $Ro$.  A Spearman's rank correlation test for
the detected sources gives $\rho\approx -0.72$ with a null hypothesis
(no correlation) probability of only $\approx 7.0\times 10^{-8}$.  A
linear regression fit indicates an overall trend of $L_{\rm rad}/R_*^2
\propto Ro^{-0.5}$.

\section{Implications for Magnetic Dynamo Models}
\label{sec:disc}

The single trend of radio activity and surface flux as a function of
Rossby number for G--L dwarfs, the overall increase in radio activity
with rotation velocity, and the enhanced fraction of radio emitters at
$v{\rm sin}i\gtrsim 25$ km s$^{-1}$ indicate that rotation continues
to play a role in the dynamo mechanism of ultracool dwarfs.  These
trends are at odds with observations in X-rays and H$\alpha$, which
point to a breakdown in the relation between activity and rotation.
This suggests that the reduced activity levels in X-rays and H$\alpha$
are due to external effects, such as the increased neutrality of the
atmospheres, a reduction in the efficiency of bulk coronal heating, or
centrifugal stripping, rather than to a substantial decrease in the
dynamo efficiency.  In this context, the radio observations provide
strong support to Zeeman measurements that point to the continued
presence of $\sim 1-3$ kG fields in some late-M dwarfs
\citep{mdp+10,rb10}.  However, ZDI measurements suggest a breakdown in
the correlation between stellar parameters (e.g., rotation) and the
field strength for $Ro\lesssim 0.1$ such that some rapid rotators have
weak fields, while others have substantial fields \citep{mdp+10}.  The
evidence from radio observations extends to faster rotation velocities
than the limit of $v{\rm sin}i\lesssim 20$ km s$^{-1}$ for Zeeman
measurements, and may be indicative of a similar trend, namely rapid
rotators are more likely to produce radio emission, and to follow the
same trend with respect to Rossby number of G--K stars, but there is a
large fraction of rapid rotators with no detectable radio emission.

With the exception of the purely turbulent dynamo model of
\citet{ddr93}, all other published models of fully-convective dynamos
predict some level of relation between the magnetic energy density and
rotation, at least up to a saturation level \citep{ck06,dsb06,bro08}.
In particular, \citet{ck06} find that for in an $\alpha^2$ dynamo the
resulting field strength depends on rotation up to a saturation value,
and it may indeed dominate in the fastest rotators leading to the
activity-rotation saturation observed in X-rays and H$\alpha$.
Similarly, \citet{dsb06} find that on the large scales the magnetic
energy increases with rotation (up to some saturation value), while
small-scale fields are nearly independent of rotation.  \citet{bro08}
also finds that faster rotators produce stronger fields, and that
rapid rotation leads to suppression of differential rotation.

The dynamo models also predict that rotation will affect the field
topology, but there is little agreement about the resulting field
configurations.  Predictions range from dominant axisymmetric fields
\citep{bro08}, with a primary quadrupolar component \citep{dsb06}, to
large-scale, non-axisymmetric, high multipole order fields
\citep{ck06}.  As in the case of field strength, field topology
measurements with the ZDI technique suggest that in late-M dwarfs
there is no clear correlation between stellar parameters (e.g.,
rotation) and the field topology.  Since a substantial fraction of the
radio emitters produce simple rotationally modulated emission
indicative of a dipolar field topology
\citep{brr+05,had+06,hbl+07,brp+09,mbi+11}, it is possible that
magnetic topology rather than field strength is the key to the change
in the nature of magnetic activity among ultracool dwarfs, and that
this is the main parameter that is correlated with rotation.  A clear
test of this possibility is to observe all radio emitters for at least
a few rotation periods to test for periodicity.  For the objects with
$v{\rm sin}i\lesssim 20$ km s$^{-1}$ this will require $\sim 40$ hr
per source.

\section{Conclusions}
\label{sec:conc}

We presented new observations of a large sample of M and L dwarfs with
measured rotation velocities aimed at addressing the radio
activity-rotation relation in fully convective objects.  This survey
triples the number of ultracool dwarfs with measured rotation
velocities and radio observations.  As part of this survey we also
discovered three new radio active ultracool dwarfs, of which one
(2M1314+1320) exhibits periodic radio emission \citep{mbi+11}.
Combining our observations with objects from the literature we find
the following key results:

\begin{itemize}

\item In the M0--M6 dwarfs we find a saturation-type relation between
rotation period and radio activity, similar to the one seen in
H$\alpha$ and X-ray, and reaching saturation at a relatively low
rotation velocity of about 5 km s$^{-1}$.

\item Unlike the rapid decline in X-ray and H$\alpha$ activity in
ultracool dwarfs, even for the most rapid rotators, the radio
luminosity remains unchanged as a function of rotation velocity and
spectral type, at least to spectral type of about L4.  The ratio of
radio to bolometric luminosity increases with later spectral type,
well beyond the saturation value of M0--M6 dwarfs.  However, as in the
case of X-ray and H$\alpha$ activity we find an increased scatter in
$L_{\rm rad}$ and $L_{\rm rad}/L_{\rm bol}$ for ultracool dwarfs.

\item In the regime of fastest rotation ($v\sin i\gtrsim 20$ km
s$^{-1}$), there are fewer objects with low radio luminosity and a
higher fraction of detected objects.  This is contrary to the apparent
X-ray and H$\alpha$ super-saturation in these fast rotators.

\item The ratio of radio to bolometric luminosity and the radio
surface flux increase as a function of decreasing Rossby number with a
single trend for $Ro\sim 0.1-10^{-3}$ and spectral types G--L.  This
is in direct contrast to the saturated X-ray/H$\alpha$ activity-$Ro$
relation in G--L dwarfs, and the X-ray/H$\alpha$ super-saturation in
ultracool dwarfs.

\end{itemize}

Our most basic conclusion from these observations is that rotation
continues to play a role in the magnetic activity of ultracool dwarfs,
and hence in the underlying dynamo mechanism.  It is not possible at
the present to determine whether rotation mainly influences the field
strength or its topology, since both may affect the detectability of
radio emission.  A clear test is long-term monitoring of the radio
emitters to check for periodic modulation, which will allow us to
reconstruct the field configuration \citep{brp+09,mbi+11}.  The
ability of radio observations to trace the presence of magnetic fields
in the most rapid rotators ($v{\rm sin}i\gtrsim 20$ km s$^{-1}$) is
particularly important in light of the inability of Zeeman
measurements to probe this regime.  We are clearly able to study the
role of rotation down to $Ro\sim 10^{-3}$, while the Zeeman techniques
are sensitive only to $Ro\gtrsim 10^{-2}$ \citep{mdp+10,rb10}.

While rotation clearly plays a role in radio activity, there are rapid
rotators with no detectable radio emission, suggesting that more than
one dynamo mechanism could be operating in ultracool dwarfs, or that
the dynamo may lead to significantly different strengths/topologies.
This is similar to the conclusion of \citet{mdp+10} from ZDI
measurements of mid- and late-M dwarfs.  In particular, it is possible
that some rapid rotators are dominated by a dynamo that leads to a
large-scale, low multipole order field that is more likely to result
in detectable radio emission (particularly simple periodic radio
emission).  The long-term variability of at least some ultracool
dwarfs (\S\ref{sec:rad}; \citealt{adh+08,bbf+10}) may be indicative of
an episodic switch between the dynamo states.

Since most of the radio detections of ultracool dwarfs to date are
close to the sensitivity limit of the VLA, future studies of
individual objects and trends such as the activity-rotation relation
will greatly benefit from the order of magnitude increase in
sensitivity afforded by the now-operational EVLA.  Any constraints on
the convective dynamo mechanism, atmospheric coupling of the magnetic
field, or bulk coronal densities must take results from radio activity
studies into account, particularly for the fastest rotators.  In
addition, theoretical dynamo models should explore the range of
stellar parameters and rotation rates that are directly probed by
radio observations, extending down to at least $Ro\sim 10^{-3}$.

\acknowledgements E.B.~acknowledges support for this work from the
National Science Foundation through Grant AST-1008361.  A.R.~received
research funding from the DFG as an Emmy Noether fellow (RE 1664/4-1).
This work has made use of the SIMBAD database, operated at CDS,
Strasbourg, France.


\clearpage
\LongTables
\begin{deluxetable}{llcccccccll}
\tablecolumns{11}
\tabletypesize{\scriptsize}
\tabcolsep0.08in\footnotesize
\tablewidth{0pc}
\tablecaption{Results from Our Radio Survey of M and L Dwarfs
\label{tab:obsn}}
\tablehead {
\colhead {2MASS Number}         &
\colhead {Other Name}           &
\colhead {Sp.T.}                &
\colhead {$J$}                  &
\colhead {$K$}                  &
\colhead {$d$}                  &
\colhead {$v{\rm sin}i$}        &
\colhead {$L_{\rm bol}$}        &
\colhead {$L_{\rm H\alpha}/L_{\rm bol}$}   &       
\colhead {$F_{\nu}$}            &
\colhead {$\nu L_\nu/L_{\rm bol}$} \\ 
\colhead {}                     &
\colhead {}                     &
\colhead {}                     &
\colhead {(mag)}                &
\colhead {(mag)}                &
\colhead {(pc)}                 &
\colhead {(km s$^{-1}$)}        &
\colhead {($L_\odot$)}          &
\colhead {}                     &
\colhead {($\mu$Jy)}            &
\colhead {}                     
}
\startdata
%
$0318238\!-\!010018$    & SDSS-MEB-1   & M4.0 &  15.4 & 14.62 & 375 & \nod & $-1.81$ &   \nod  & $<78  $ & $<-5.74$ \\
$0629234\!-\!024850$A   & GJ 234 A     & M4.5 &  6.38 &  5.49 &   4 &    6 & $-2.89$ & $-3.98$ & $<81  $ & $<-8.50$ \\
$1406493\!-\!301828$    & LHS 2859     & M5.0 & 11.36 & 10.37 &  19 & \nod & $-2.84$ &   \nod  & $<81  $ & $<-7.27$ \\
$2043192\!+\!552053$    & GJ 802 A     & M5.0 &  9.56 &  8.75 &   9 &  6.4 & $-2.23$ &   \nod  & $<93  $ & $<-7.03$ \\
$0004575\!-\!170937$    &              & M5.5 & 11.00 & 10.08 &  16 & \nod & $-3.31$ & $-3.81$ & $<78  $ & $<-6.98$ \\
$1610584\!-\!063132$    & LP 684-33    & M5.5 & 11.35 & 10.37 &  18 & \nod & $-2.91$ &   \nod  & $<96  $ & $<-7.19$ \\
$2132297\!-\!051158$    & LP 698-2     & M5.5 & 11.42 & 10.38 &  19 & \nod & $-2.91$ & $-5.06$ & $<72  $ & $<-7.29$ \\
$2151270\!-\!012713$    & LP 638-50    & M5.5 & 11.28 & 10.39 &  19 & \nod & $-2.81$ &   \nod  & $<63  $ & $<-7.43$ \\
$2205357\!-\!110428$    & LP 759-25    & M5.5 & 11.66 & 10.72 &  19 &   13 & $-2.98$ & $-4.20$ & $<84  $ & $<-7.15$ \\
$1236153\!-\!310646$    &              & M5.5 & 11.78 & 10.81 &  19 & \nod & $-3.00$ &   \nod  & $<111 $ & $<-6.96$ \\
$0013466\!-\!045736$    & LHS 1042     & M6.0 & 11.46 & 10.48 &  17 & \nod & $-3.01$ &   \nod  & $<81  $ & $<-7.23$ \\
$0024441\!-\!270824$    & GJ 2005 A    & M6.0 &  9.25 &  8.24 &  19 &    9 & $-3.25$ & $-4.62$ & $<96  $ & $<-6.50$ \\
$1236396\!-\!172216$    &              & M6.0 & 11.77 & 10.63 &  19 & \nod & $-3.02$ &   \nod  & $<69  $ & $<-7.16$ \\
$1346460\!-\!314925$    & LP 911-56    & M6.0 & 10.98 & 10.04 &  14 & \nod & $-2.96$ &   \nod  & $<84  $ & $<-7.41$ \\
$1432085\!+\!081131$    & LHS 2935     & M6.0 & 10.11 &  9.17 &   9 & \nod & $-2.96$ &   \nod  & $<81  $ & $<-7.79$ \\
$1552446\!-\!262313$    & LHS 5303     & M6.0 & 10.37 &  9.30 &  11 & \nod & $-2.98$ &   \nod  & $<78  $ & $<-7.67$ \\
$1614252\!-\!025100$    &              & M6.0 & 11.30 & 10.28 &  18 & \nod & $-2.87$ & $-4.19$ & $<90  $ & $<-7.23$ \\
$2049527\!-\!171608$    &              & M6.0 & 11.81 & 10.81 &  19 & \nod & $-3.02$ &   \nod  & $<60  $ & $<-7.22$ \\
$0518113\!-\!310153$    &              & M6.5 & 11.88 & 10.90 &  20 & \nod & $-3.04$ &   \nod  & $181\pm 27$ & $-6.73$  \\
$0931223\!-\!171742$    & LP 788-1     & M6.5 & 11.07 & 10.07 &  13 & \nod & $-3.09$ &   \nod  & $<54  $ & $<-7.55$ \\
$1516407\!+\!391048$    & LP 222-65    & M6.5 & 10.80 &  9.81 &  12 & \nod & $-3.23$ &   \nod  & $<81  $ & $<-7.30$ \\
$1606339\!+\!405421$    & LHS 3154     & M6.5 & 11.05 & 10.07 &  12 & \nod & $-3.34$ &   \nod  & $<117 $ & $<-7.02$ \\
$1646315\!+\!343455$    & LHS 3241     & M6.5 & 10.53 &  9.61 &  11 & \nod & $-3.15$ &   \nod  & $<84  $ & $<-7.46$ \\
$0535218\!-\!054608$AB  &              & M6.5 & 14.65 & 13.47 & 156 & \nod & $-2.35$ &   \nod  & $<93  $ & $<-5.88$ \\
$0711113\!+\!432959$    & LHS 1901     & M6.5 &  9.98 &  9.13 &  13 & \nod & $-2.94$ &   \nod  & $<117 $ & $<-7.35$ \\
$0741068\!+\!173845$    & LHS 1937     & M7.0 & 12.01 & 10.94 &  18 &   10 & $-3.17$ & $-4.10$ & $<75  $ & $<-7.03$ \\
$0818580\!+\!233352$    &              & M7.0 & 12.18 & 11.15 &  19 &  4.5 & $-3.19$ & $-4.11$ & $<78  $ & $<-6.94$ \\
$0952219\!-\!192431$    &              & M7.0 & 11.87 & 10.87 &  30 &    6 & $-3.08$ & $-3.94$ & $<69  $ & $<-6.43$ \\
                        &              &      &       &       &     &      &         &         & $233\pm 15$ & $ -6.20$ \\
$1048126\!-\!112009$    & GJ 3622      & M7.0 &  8.86 &  7.93 &   5 & $<3$ & $-3.09$ & $-4.63$ & $<96  $ & $<-8.20$ \\
$1141440\!-\!223215$    &              & M7.0 & 12.63 & 11.57 &  22 &   10 & $-3.25$ & $-4.90$ & $<108 $ & $<-6.62$ \\
$1314203\!+\!132001$A   &              & M7.0 &  9.75 &  8.79 &  16 &   45 & $-3.17$ & $-3.97$ & $1156\pm 15$ & $ -5.92$ \\
$1354087\!+\!084608$    &              & M7.0 & 12.19 & 11.16 &  17 & \nod & $-3.79$ &   \nod  & $<105 $ & $<-6.31$ \\
$1356414\!+\!434258$    & LP 220-13    & M7.0 & 11.71 & 10.65 &  16 &   14 & $-3.59$ & $-3.92$ & $<99  $ & $<-6.61$ \\
$1534570\!-\!141848$    & 2MUCD 11346  & M7.0 & 11.38 & 10.31 &  11 &   10 & $-3.34$ & $-4.01$ & $<87  $ & $<-7.21$  \\
$2337383\!-\!125027$    & LP 763-3     & M7.0 & 11.46 & 10.45 &  19 & \nod & $-2.89$ & $-3.50$ & $<84  $ & $<-7.20$ \\
$0351000\!-\!005244$    & GJ 3252      & M7.5 & 11.30 & 10.23 &  15 &  6.5 & $-3.06$ & $-4.16$ & $<123 $ & $<-7.10$ \\
$1006319\!-\!165326$    & LP 789-23    & M7.5 & 12.04 & 10.99 &  16 &   16 & $-3.28$ & $-4.22$ & $<87  $ & $<-6.96$ \\
$1155429\!-\!222458$    & LP 851-346   & M7.5 & 10.94 &  9.88 &  10 &   33 & $-3.30$ & $-4.58$ & $<90  $ & $<-7.36$ \\
$1246517\!+\!314811$    & LHS 2632     & M7.5 & 12.23 & 11.21 &  18 &  7.3 & $-3.25$ & $-5.27$ & $<90  $ & $<-6.86$ \\
$1250526\!-\!212113$    &              & M7.5 & 11.16 & 10.13 &  11 & \nod & $-3.25$ &   \nod  & $<72  $ & $<-7.39$ \\
$1253124\!+\!403403$    & LP 218-8     & M7.5 & 12.19 & 11.16 &  17 &    9 & $-3.29$ & $-4.27$ & $<78  $ & $<-6.94$ \\
$1332244\!-\!044112$    &              & M7.5 & 12.37 & 11.28 &  21 &    9 & $-3.18$ & $-4.37$ & $<60  $ & $<-6.97$ \\
$1507277\!-\!200043$    &              & M7.5 & 11.71 & 10.66 &  14 &   64 & $-3.61$ & $-4.47$ & $<96  $ & $<-6.69$ \\
$1546054\!+\!374946$    &              & M7.5 & 12.44 & 11.41 &  20 &   10 & $-3.25$ & $-3.98$ & $<84  $ & $<-7.04$ \\
$1757154\!+\!704201$    & LP 44-162    & M7.5 & 11.45 & 10.40 &  12 &   33 & $-3.30$ & $-5.01$ & $<117 $ & $<-7.05$ \\
$2331217\!-\!274950$    &              & M7.5 & 11.30 & 10.23 &  15 &    9 & $-3.06$ & $-4.03$ & $<72  $ & $<-7.32$ \\
$0027559\!+\!221932$    & LP 349-25 B  & M8.0 & 10.61 &  9.57 &  10 &   56 & $-3.12$ & $-4.53$ & $323\pm 14$ & $ -6.95$ \\
$0248410\!-\!165121$    &              & M8.0 & 12.55 & 11.42 &  17 & $<3$ & $-3.45$ & $-4.25$ & $<81  $ & $<-6.77$ \\
$0320596\!+\!185423$    & LP 412-31    & M8.0 & 11.76 & 10.64 &  15 &   15 & $-3.26$ & $-3.87$ & $<81  $ & $<-7.08$ \\
$0544115\!-\!243301$    &              & M8.0 & 12.53 & 11.46 &  19 & $<3$ & $-3.33$ & $-4.12$ & $<63  $ & $<-6.90$ \\
$0629235\!-\!024851$B   & GJ 234 B     & M8.0 &  8.38 &  7.33 &   4 & \nod & $-3.00$ &   \nod  & $<81  $ & $<-8.45$ \\
$1016347\!+\!275149$    & LHS 2243     & M8.0 & 11.99 & 10.96 &  14 & $<3 $& $-3.38$ & $-3.87$ & $<84  $ & $<-6.99$  \\
$1024099\!+\!181553$    & 2MUCD 10906  & M8.0 & 12.28 & 11.24 &  16 &    5 & $-3.38$ & $-4.84$ & $<87  $ & $<-6.86$  \\
$1309218\!-\!233035$    &              & M8.0 & 11.79 & 10.67 &  13 &    7 & $-3.63$ & $-4.35$ & $<93  $ & $<-6.74$ \\
$1428041\!+\!135613$    & LHS 2919     & M8.0 & 11.01 & 10.03 &  10 & \nod & $-3.37$ &   \nod  & $<90  $ & $<-7.25$ \\
$1440229\!+\!133923$    &              & M8.0 & 12.40 & 11.34 &  18 & $<3$ & $-3.33$ & $-4.60$ & $<75  $ & $<-6.87$ \\
$1444171\!+\!300214$    & LP 326-21    & M8.0 & 11.67 & 10.62 &  13 & \nod & $-3.61$ &   \nod  & $<81  $ & $<-6.87$ \\
$1843221\!+\!404021$    & GJ 4073      & M8.0 & 11.31 & 10.31 &  14 &    5 & $-3.09$ & $-4.11$ & $<96  $ & $<-7.21$  \\
$2206227\!-\!204706$    &              & M8.0 & 12.37 & 11.32 &  27 &   24 & $-2.95$ & $-4.54$ & $<84  $ & $<-6.83$ \\
$2349489\!+\!122438$    & LP 523-55    & M8.0 & 12.60 & 11.56 &  20 &    4 & $-3.31$ & $-4.61$ & $<60  $ & $<-6.89$ \\
$2351504\!-\!253736$A   &              & M8.0 & 12.47 & 11.27 &  18 &   36 & $-3.36$ & $-4.61$ & $<69  $ & $<-6.88$ \\
$1121492\!-\!131308$    & GJ 3655      & M8.5 & 11.93 & 10.74 &  12 &   27 & $-3.68$ & $-3.87$ & $<102 $ & $<-6.74$ \\
$1124048\!+\!380805$    &              & M8.5 & 12.71 & 11.57 &  19 &  7.5 & $-3.41$ & $-5.16$ & $<66  $ & $<-6.80$ \\
$1403223\!+\!300754$    &              & M8.5 & 12.68 & 11.60 &  19 &   10 & $-3.39$ & $-4.49$ & $<60  $ & $<-6.86$ \\
$2353594\!-\!083331$    &              & M8.5 & 13.03 & 11.93 &  22 &  4.5 & $-3.41$ & $-4.42$ & $<69  $ & $<-6.64$ \\
$0443376\!+\!000205$    &              & M9.0 & 12.51 & 11.22 &  16 & 13.5 & $-3.47$ & $-5.00$ & $<54  $ & $<-6.99$ \\
$1224522\!-\!123835$    &              & M9.0 & 12.57 & 11.35 &  17 &    7 & $-3.94$ & $-4.52$ & $<102 $ & $<-6.20$ \\
$1411213\!-\!211950$    &              & M9.0 & 12.44 & 11.33 &  16 &   44 & $-3.93$ & $-4.93$ & $<93  $ & $<-6.29$ \\
$1428432\!+\!331039$    & LHS 2924     & M9.0 & 11.99 & 10.74 &  11 &   11 & $-3.59$ & $-5.14$ & $<84  $ & $<-6.98$ \\
$1707234\!-\!055824$    & 2MUCD 20701  & M9.0 & 12.25 & 10.90 &  17 & \nod & $-3.31$ &    \nod & $<81  $ & $<-6.92$  \\
$2200020\!-\!303832$AB  &              & M9.0 & 13.44 & 12.20 &  35 &   17 & $-3.17$ & $-5.03$ & $<78  $ & $<-6.44$ \\
$0024246\!-\!015819$    & BRI B0021-02 & M9.5 & 11.99 & 10.54 &  12 &   33 & $-3.45$ & $-6.12$ & $<60  $ & $<-7.09$ \\
$1438082\!+\!640836$    &              & M9.5 & 12.99 & 11.65 &  18 &   12 & $-4.08$ & $-4.77$ & $<105 $ & $<-5.96$ \\
$2237325\!+\!392239$    & G216-7B      & M9.5 & 13.34 & 12.18 &  19 & \nod & $-3.66$ & $-5.02$ & $<81  $ & $<-6.46$  \\
$0314034\!+\!160305$    &              & L0.0 & 12.53 & 11.24 &  14 &   19 & $-3.59$ & $-4.69$ & $<108 $ & $<-6.66$ \\
$1159385\!+\!005726$    &              & L0.0 & 14.08 & 12.81 &  30 &   71 & $-3.57$ & $-5.06$ & $<54  $ & $<-6.35$ \\
$1221277\!+\!025719$    &              & L0.0 & 13.17 & 11.95 &  19 &   25 & $-3.59$ & $-4.88$ & $<78  $ & $<-6.54$ \\
$1731297\!+\!272123$    &              & L0.0 & 12.09 & 10.91 &  12 &   15 & $-3.56$ & $-4.80$ & $<69  $ & $<-7.03$ \\
$1854459\!+\!842947$    &              & L0.0 & 13.66 & 12.47 &  23 &    7 & $-3.62$ & $-4.73$ & $<87  $ & $<-6.29$ \\
$1412244\!+\!163311$    &              & L0.5 & 13.89 & 12.52 &  25 &   19 & $-3.61$ & $-5.50$ & $<69  $ & $<-6.33$ \\
$1441371\!-\!094559$    &              & L0.5 & 14.02 & 12.66 &  28 &   23 & $-3.59$ & $-5.48$ & $<84  $ & $<-6.20$ \\
$2351504\!-\!253736$B   &              & L0.5 & 12.47 & 11.27 &  18 &   41 & $-3.36$ & $-5.22$ & $<69  $ & $<-6.87$ \\
$0235599\!-\!233120$    &              & L1.0 & 13.67 & 12.19 &  21 &   13 & $-3.63$ & $-6.44$ & $<99  $ & $<-6.30$ \\
$1045240\!-\!014957$    &              & L1.0 & 13.16 & 11.78 &  17 & $<3$ & $-3.65$ & $-6.44$ & $<57  $ & $<-6.71$ \\
$1048428\!+\!011158$    &              & L1.0 & 12.92 & 11.62 &  15 &   17 & $-3.69$ & $-5.71$ & $<21  $ & $<-7.22$ \\
$1439283\!+\!192914$    &              & L1.0 & 12.76 & 11.55 &  14 &   11 & $-3.67$ & $-5.20$ & $<78  $ & $<-6.71$ \\
$1555157\!-\!095605$    &	           & L1.0 & 12.56 & 11.44 &  13 &   11 & $-3.68$ & $-5.35$ & $<84  $ & $<-6.75$ \\
$1145571\!+\!231729$    & GL Leo       & L1.5 & 15.39 & 13.95 &  44 &   14 & $-3.70$ & $-5.27$ & $<90  $ & $<-5.65$ \\
$1334062\!+\!194035$    &              & L1.5 & 15.48 & 14.00 &  46 &   30 & $-3.68$ & $-6.53$ & $<60  $ & $<-5.74$ \\
$1645221\!-\!131951$    &              & L1.5 & 12.45 & 11.15 &  12 &    9 & $-3.69$ & $-5.66$ & $<108 $ & $<-6.71$ \\
$1807159\!+\!501531$    & 2MUCD 11756  & L1.5 & 12.93 & 11.62 &  15 &   76 & $-3.71$ & $-5.26$ & $<84  $ & $<-6.62$  \\
$0828341\!-\!130919$    &              & L2.0 & 12.80 & 11.30 &  14 &   33 & $-3.64$ & $-6.63$ & $<66  $ & $<-6.83$ \\
$0921141\!-\!210444$    & DENIS-092114 & L2.0 & 12.78 & 11.69 &  12 &   15 & $-3.83$ & $<-6.42$& $<75  $ & $<-6.72$  \\
$1029216\!+\!162652$    &              & L2.5 & 14.29 & 12.62 &  23 &   29 & $-3.72$ & $-5.76$ & $<33  $ & $<-6.64$ \\
$1047310\!-\!181557$    &              & L2.5 & 14.20 & 12.89 &  22 &   15 & $-3.86$ & $-5.99$ & $<63  $ & $<-6.23$ \\
$0913032\!+\!184150$    &              & L3.0 & 15.97 & 14.28 &  46 &   34 & $-3.77$ & $-6.86$ & $<102 $ & $<-5.47$ \\
$1203581\!+\!001550$    &              & L3.0 & 14.01 & 12.48 &  19 &   39 & $-3.84$ & $-6.02$ & $<63  $ & $<-6.39$ \\
$1506544\!+\!132106$    &	           & L3.0 & 13.37 & 11.74 &  14 &   20 & $-3.80$ & $-6.32$ & $<78  $ & $<-6.60$  \\
$1615441\!+\!355900$    &	           & L3.0 & 14.54 & 12.94 &  24 &   13 & $-3.82$ & $<-5.98$& $<75  $ & $<-6.13$  \\
$1707234\!-\!055824$    & 2MUCD 20701  & L3.0 & 13.96 & 12.25 &  17 & \nod & $-3.80$ &   \nod  & $<81  $ & $<-6.42$  \\
$0700366\!+\!315726$A   &              & L3.5 & 13.23 & 11.62 &  12 &   41 & $-3.88$ & $-6.04$ & $<78  $ & $<-6.39$ 
\enddata
\tablecomments{Properties of the M and L dwarfs observed in this
paper.  The columns are (left to right): (i) 2MASS number; (ii) other
name; (iii) spectral type; (iv) $J$-band magnitude; (v) $K$-band
magnitude; (vi) distance from parallax or photometric estimate (from
SIMBAD, \citealt{fbc+09}, and \citealt{crl+03}); (vii) projected
rotation velocity (from \citealt{mb03,rkl+02,rb08,jrj+09,rb10});
(viii) bolometric luminosity; (ix) H$\alpha$ activity (from
\citealt{dfp+98,mb03,rb08,rb10}); (x) radio density flux; and (xi)
ratio of radio to bolometric luminosity.}
\end{deluxetable}

\clearpage
\begin{deluxetable}{llcccccccllc}
\tablecolumns{12}
\tabletypesize{\scriptsize}
\tabcolsep0.08in\footnotesize
\tablewidth{0pc}
\tablecaption{Results from the Literature
\label{tab:obso}}
\tablehead {
\colhead {2MASS Number}         &
\colhead {Other Name}           &
\colhead {Sp.T.}                &
\colhead {$J$}                  &
\colhead {$K$}                  &
\colhead {$d$}                  &
\colhead {$v{\rm sin}i$}        &
\colhead {$L_{\rm bol}$}        &
\colhead {$L_{\rm H\alpha}/L_{\rm bol}$}   &       
\colhead {$F_{\nu}$}            &
\colhead {$\nu L_\nu/L_{\rm bol}$}   &   
\colhead {Ref}                  \\
\colhead {}                     &
\colhead {}                     &
\colhead {}                     &
\colhead {(mag)}                &
\colhead {(mag)}                &
\colhead {(pc)}                 &
\colhead {(km s$^{-1}$)}        &
\colhead {($L_\odot$)}          &
\colhead {}                     &
\colhead {($\mu$Jy)}            &
\colhead {}                     &
\colhead {}                     
}
\startdata
%
$1120052\!+\!655047$  & GJ 424      & M0.0 &  6.31 &  5.53 &   9 &     \nod & $<-1.71$& $<-5.0$ & $<240$ & $<-8.33$ & 1\\      
$1300466\!+\!122232$  & GJ 494B     & M0.5 &  6.44 &  5.58 &  11 &       10 & $-1.63$ & \nod    &  $340$ & $ -8.06$ & 1\\        
$0042482\!+\!353255$  & GL29.1      & M1.0 &  7.16 &  6.32 &  24 &     \nod & $-1.27$ & \nod    & $<150$ & $<-8.14$ & 1\\       
$0610346\!-\!215152$  & GJ 229A     & M1.0 &  5.10 &  4.17 &   6 &        1 & $-1.74$ & \nod    & $<290$ & $<-8.61$ & 1\\       
$0102389\!+\!622042$  & GL49        & M1.5 &  6.23 &  5.37 &  10 & $<3.4$   & $-1.66$ & \nod    & $<370$ & $<-8.10$ & 1\\       
$0018225\!+\!440122$  & GL15A       & M2.0 &  5.25 &  4.02 &   4 &      2.9 & $-2.37$ & \nod    & $<220$ & $<-8.52$ & 1\\       
$1103202\!+\!355811$  & GJ 411      & M2.0 &  4.20 &  3.25 &   3 &  $<2.9$  & $-2.10$ & $<-5$   & $<300$ & $<-9.49$ & 1\\      
$1105290\!+\!433135$  & GJ 412A     & M2.0 &  5.54 &  4.77 &   5 &  $<3.0$  & $-1.95$ & $<-5$   & $<220$ & $<-8.68$ & 1\\     
$2238455\!-\!203716$  & GJ 867A     & M2.0 &  5.67 &  4.80 &   9 &     \nod & $-1.58$ & \nod    & $<600$ & $<-8.11$ & 1\\       
$2349125\!+\!022403$  & GJ 908      & M2.0 &  5.83 &  5.04 &   6 & $<3.1$   & $-1.90$ & $<-5$   & $<200$ & $<-8.59$ & 1\\      
$0533448\!+\!015643$  & GJ 207.1    & M2.5 &  7.76 &  6.86 &  17 &    10    & $-1.86$ & \nod    & $<150$ & $<-7.85$ & 1\\        
$1332446\!+\!164839$  & GJ 516A     & M2.5 &  7.64 &  6.83 &  20 &     \nod & $-1.25$ & \nod    & $<320$ & $<-8.10$ & 1\\        
$1454292\!+\!160603$  & GJ 569AB    & M2.5 &  6.63 &  5.77 &  10 &   $<2.5$ & $-1.84$ & \nod    & $<390$ & $<-7.92$ & 1\\        
$0032297\!+\!671404$  & GL22B       & M3.0 &  7.17 &  6.38 &  10 &     \nod & $-1.97$ & \nod    & $<390$ & $<-7.77$ & 1\\        
$1332446\!+\!164839$  & GJ 516B     & M3.0 &  7.64 &  6.83 &  14 &     \nod & $-1.91$ & \nod    & $<320$ & $<-7.65$ & 1\\        
$1655528\!-\!082010$  & GJ 644A     & M3.0 &  5.27 &  4.40 &   7 &     \nod & $-1.66$ & \nod    & $1220$ & $ -7.97$ & 1\\          
$1842466\!+\!593749$  & GL 725A     & M3.0 &  5.19 &  4.43 &   4 &     $<5$ & $-2.06$ & $<-5$   & $<180$ & $<-8.92$ & 1\\        
$1849492\!-\!235010$  & GJ 729      & M3.0 &  6.22 &  5.37 &   3 &        4 & $-2.71$ & \nod    & $<300$ & $<-8.21$ & 1\\        
$1855274\!+\!082409$  & GJ 735      & M3.0 &  6.31 &  5.43 &  12 &    $<10$ & $-1.58$ & \nod    & $450\pm 150$ & $ -7.98$ & 1\\       
$0004364\!-\!404402$  & GJ 1001A    & M3.5 &  8.60 &  7.74 &  10 &     \nod & $-2.66$ & \nod    & $<45$  & $<-8.31$ & 2\\        
$0032297\!+\!671408$  & GL22A       & M3.5 &  6.84 &  6.04 &  10 &     \nod & $-1.85$ & \nod    & $<390$ & $<-7.89$ & 1\\        
$0532146\!+\!094915$  & GJ 206      & M3.5 &  7.42 &  6.56 &  13 &       10 & $-1.93$ & \nod    & $1500$ & $ -7.56$ & 1\\          
$1736259\!+\!682022$  & GJ 687      & M3.5 &  5.34 &  4.55 &   5 &     $<5$ & $-1.94$ & $<-5$   & $300$  & $ -9.15$ & 1\\         
$1842468\!+\!593737$  & GL 725B     & M3.5 &  5.72 &  5.00 &   4 &     $<7$ & $-2.25$ & $<-5$   & $<180$ & $<-8.74$ & 1\\       
$1916552\!+\!051008$  & GJ 752A     & M3.5 &  5.58 &  4.67 &   6 &   $<2.6$ & $-1.90$ & $<-5$   & $ 290$ & $ -8.44$ & 1\\        
$2238453\!-\!203651$  & GJ 867B     & M3.5 &  7.34 &  6.49 &   9 &     \nod & $-2.23$ & \nod    & $ 810$ & $ -7.33$ & 1\\          
$2331520\!+\!195614$  & GJ 896A     & M3.5 &  6.16 &  5.33 &   6 &       10 & $-2.02$ & \nod    & $ 570$ & $ -7.97$ & 1\\          
$0112305\!-\!165957$  & GL54.1      & M4.0 &  7.26 &  6.42 &   4 &   $<2.5$ & $-2.92$ & \nod    & $<390$ & $<-7.69$ & 1\\        
$0139011\!-\!175701$  & GL65A       & M4.0 &  6.28 &  5.34 &   3 &     29.4 & $-2.89$ & \nod    & $400$  & $ -7.76$ & 1\\          
                      &             &      &       &       &     &          & $     $ &         & $4000$ & $ -6.99$ & 1\\          
$0431114\!+\!585837$  & GJ 169.1AB  & M4.0 &  6.62 &  5.72 &   6 &      1.9 & $-2.35$ & $-5.16$ & $<150$ & $<-8.31$ & 1\\        
$1147444\!+\!004816$  & GJ 447      & M4.0 &  6.51 &  5.65 &   3 &     $<2$ & $-2.73$ & $<-5$   & $<210$ & $<-8.24$ & 1\\       
$1233163\!+\!090126$  & GJ 473A     & M4.0 &  6.88 &  6.04 &   4 &     \nod&  $-2.75$ & \nod    & $ 200$ & $ -7.92$ & 1\\          
$2227595\!+\!574145$  & GJ 860B     & M4.0 &  5.58 &  4.78 &   4 &      4.7 & $-2.15$ & $-4.11$ & $1283$ & $ -7.35$ & 1\\         
$0200127\!+\!130311$  & GL83.1      & M4.5 &  7.51 &  6.65 &   5 &      3.8 & $-2.87$ & $-4.35$ & $<260$ & $<-7.75$ & 1\\        
$0415217\!-\!073917$  & GJ 166C     & M4.5 &  6.75 &  5.96 &   5 &        5 & $-2.41$ & $-3.95$ & $<270$ & $<-8.09$ & 1\\        
$0629234\!-\!024850$  & GJ 234A     & M4.5 &  6.38 &  5.49 &   4 &        6 & $-2.89$ & $-3.98$ & $ 420$ & $ -8.45$ & 1\\
$0710018\!+\!383145$  & GJ 268      & M4.5 &  6.73 &  5.85 &   6 &     \nod & $-2.27$ & \nod    & $ 330$ & $ -8.04$ & 1\\          
$0744401\!+\!033308$  & GJ 285      & M4.5 &  6.58 &  5.70 &   6 &      6.5 & $-2.27$ & $-3.18$ & $ 300$ & $ -7.94$ & 1\\          
                      &             &      &       &       &     &          & $     $ &         & $2000$ & $ -7.22$ & 1\\          
$1019363\!+\!195212$  & GJ 388      & M4.5 &  5.45 &  4.59 &   5 &      2.7 & $-2.01$ & $-3.77$ & $ 200$ & $ -8.68$ & 1\\          
$1300335\!+\!054108$  & GJ 493.1    & M4.5 &  8.55 &  7.66 &   8 &     16.8 & $-2.80$ & $-3.96$ & $1280$ & $ -7.09$ & 1\\          
$1634204\!+\!570943$  & GJ 630.1A   & M4.5 &  8.50 &  7.80 &  15 &     27.5 & $-2.10$ & \nod    & $<530$ & $<-7.18$ & 1\\        
$1719529\!+\!263002$  & GJ 669B     & M4.5 &  8.23 &  7.35 &  12 &    $<10$ & $-2.32$ & \nod    & $ 510$ & $ -7.15$ & 1\\     
$2029483\!+\!094120$  & GJ 791.2    & M4.5 &  8.23 &  7.31 &  10 &     \nod & $-2.74$ & $-3.84$ & $<350$ & $<-7.10$ & 1\\        
$2246498\!+\!442003$  & GJ 873A     & M4.5 &  6.11 &  5.30 &   5 &      6.9 & $-2.17$ & $-3.70$ & $<300$ & $<-8.28$ & 1\\        
$0103197\!+\!622155$  & GL51        & M5.0 &  8.61 &  7.72 &  10 &     \nod & $-2.60$ & \nod    & $7280$ & $ -5.83$ & 1\\        
$2217189\!-\!084812$  & GJ 852A     & M5.0 &  9.02 &  8.17 &  10 &     \nod & $-2.97$ & \nod    & $<290$ & $<-6.87$ & 1\\        
$2217187\!-\!084818$  & GJ 852B     & M5.0 &  9.46 &  8.53 &   9 &       32 & $-2.61$ & \nod    & $<290$ & $<-7.28$ & 1\\        
$1953544\!+\!442454$  & GJ 1245A    & M5.5 &  7.79 &  6.85 &   5 &     22.5 & $-3.01$ & $-4.27$ & $<310$ & $<-7.50$ & 1\\        
$1953550\!+\!442455$  & GJ 1245B    & M5.5 &  8.28 &  7.39 &   5 &      6.8 & $-3.14$ & $-4.25$ & $<310$ & $<-7.34$ & 1\\        
$0018254\!+\!440137$  & GL15B       & M6.0 &  6.79 &  5.95 &   4 &   $<3.1$ & $-2.77$ & \nod    & $<220$ & $<-8.12$ & 1\\          
$0024441\!-\!270825$  & GJ 2005A    & M6.0 &  9.25 &  8.24 &   7 &        9 & $-3.25$ & $-4.62$ & $161\pm 15$ & $ -7.39$ & 2\\   
$0139012\!-\!175702$  & GL65B       & M6.0 &  6.28 &  5.34 &   3 &     31.5 & $-2.91$ & \nod    & $1500$ & $ -7.41$ & 1\\          
                      &             &      &       &       &     &          & $     $ &         & $3000$ & $ -7.11$ & 1\\          
$1105313\!+\!433117$  & GJ 412B     & M6.0 &  8.74 &  7.84 &   5 &      7.7 & $-3.33$ & $-3.95$ & $<220$ & $<-7.30$ & 1\\         
$1056288\!+\!070052$  & GJ 406      & M6.5 &  7.09 &  6.08 &   2 &   $<3$   & $-3.37$ & $-3.89$ & $<390$ & $<-7.63$ & 1\\         
$0435161\!-\!160657$  & LP 775-31   & M7.0 &  10.4 &  9.34 &   9 &     \nod & $-3.59$ & $-4.28$ & $<48$  & $<-7.44$ & 2\\         
$0440232\!-\!053008$  & LP 655-48   & M7.0 & 10.68 &  9.56 &  10 &     16.5 & $-3.62$ & $-3.80$ & $<39$  & $<-7.39$ & 2\\         
$0752239\!+\!161215$  & LP 423-31   & M7.0 & 10.83 &  9.82 &  11 &        9 & $-3.56$ & $-3.44$ & $<39$  & $<-7.39$ & 2\\         
$1456383\!-\!280947$  & GJ 3877     & M7.0 &  9.96 &  8.92 &   7 &        8 & $-3.29$ & $-4.02$ & $270\pm 40$ & $ -7.23$ & 3\\      
$1634216\!+\!571008$  & GJ 630.1B   & M7.0 & 14.11 & 14.14 &  16 &     \nod & $-3.13$ & \nod    & $<530$ & $<-6.05$ & 1\\         
$1655352\!-\!082340$  & VB 8        & M7.0 &  9.78 &  8.83 &   6 &        9 & $-3.21$ & \nod    & $<24$  & $<-8.37$ & 4\\         
$0148386\!-\!302439$  &             & M7.5 & 12.28 & 11.24 &  18 &       48 & $-3.67$ & $-4.35$ & $<45$  & $<-6.73$ & 2\\         
$0331302\!-\!304238$  & LP 888-18   & M7.5 & 11.37 & 10.28 &  12 &     $<3$ & $-3.70$ & $-4.07$ & $<72$  & $<-6.86$ & 2\\          
$0417374\!-\!080000$  &             & M7.5 & 12.17 & 11.05 &  17 &        7 & $-3.72$ & $-4.32$ & $<36$  & $<-6.83$ & 2\\         
$0429184\!-\!312356$  &             & M7.5 & 10.89 &  9.80 &  10 &    $<3$  & $-3.70$ & $-3.93$ & $<48$  & $<-7.23$ & 2\\          
$1521010\!+\!505323$  & NLTT 40026  & M7.5 & 12.00 & 10.92 &  16 &       40 & $-3.70$ & $-4.88$ & $<39$  & $<-6.88$ & 2\\         
$0019262\!+\!461407$  &             & M8.0 & 12.61 & 11.47 &  19 &       68 & $-3.80$ & $-4.51$ & $<33$  & $<-6.68$ & 2\\         
$0350573\!+\!181806$  & LP 413-53   & M8.0 & 12.95 & 11.76 &  23 &        4 & $-3.82$ & \nod    & $<105$ & $<-6.02$ & 2\\        
$0436103\!+\!225956$  &             & M8.0 & 13.76 & 12.19 & 140 &     \nod & $-2.62$ & \nod    & $<45$  & $<-6.02$ & 2\\         
$0517376\!-\!334902$  &             & M8.0 & 12.00 & 10.82 &  15 &        8 & $-3.82$ & $-4.42$ & $<54$  & $<-6.70$ & 2\\         
$1016347\!+\!275149$  & LHS 2243    & M8.0 & 11.95 & 10.95 &  16 &     $<3$ & $-3.65$ & $-3.87$ & $<45$  & $<-6.88$ & 5\\          
$1048146\!-\!395606$  & DENIS 1048  & M8.0 &  9.55 &  8.45 &   4 &       18 & $-3.39$ & $-5.15$ & $140\pm 40$    & $-7.83$ & 3\\     
                      &             &      &       &       &     &          & $     $ &         & $29600\pm 100$ & $-5.51$ & 3\\
$1139511\!-\!315921$  &             & M8.0 & 12.67 & 11.49 &  20 &     \nod & $-3.39$ & \nod    & $<99$  & $<-6.60$ & 3\\        
$1534570\!-\!141848$  &             & M8.0 & 11.39 & 10.31 &  11 &       10 & $-3.39$ & $-4.01$ & $<111$ & $<-7.06$ & 3\\        
$1843221\!+\!404021$  & GJ 4073     & M8.0 & 11.30 & 10.27 &  14 &        5 & $-3.51$ & $-4.11$ & $<48$  & $<-7.10$ & 2\\         
$1916576\!+\!050902$  & VB 10       & M8.0 &  9.95 &  8.81 &   6 &      6.5 & $-3.35$ & $<-5$   & $<81$  & $<-7.81$ & 4\\         
$2037071\!-\!113756$  &             & M8.0 & 12.28 & 11.26 &  17 &    $<3$  & $-3.74$ & $-5.02$ & $<33$  & $<-6.87$ & 2\\          
$0335020\!+\!234235$  &             & M8.5 & 12.26 & 11.26 &  19 &     \nod & $-3.61$ & \nod    & $<69$  & $<-6.57$ & 2\\         
$1454290\!+\!160605$  & GJ 569Ba    & M8.5 & 11.14 & 10.02 &  10 &     \nod & $-3.80$ & \nod    & $<30$  & $<-7.32$ & 2\\         
$1501081\!+\!225002$  & TVLM513-465 & M8.5 & 11.80 & 10.74 &  11 &       60 & $-3.59$ & \nod    & $190\pm 15$ & $-6.66$ & 5\\       
                      &             &      &       &       &     &          & $     $ &         & $980\pm 40$ & $-5.95$ & 5\\      
$1835379\!+\!325954$  & LSR J1835+3 & M8.5 & 10.27 &  9.15 &   6 &       44 & $-3.93$ & $-4.85$ & $525\pm 15$ & $-6.42$ & 2\\      
$0140026\!+\!270150$  &             & M8.5 & 12.49 & 11.43 &  19 &      6.5 & $-3.32$ & \nod    & $<20$  & $<-6.93$ & 6\\       
$0019457\!+\!521317$  &             & M9.0 & 12.82 & 11.62 &  19 &        9 & $-3.95$ & $-4.29$ & $<42$  & $<-6.47$ & 2\\         
$0109511\!-\!034326$  & LP 647-13   & M9.0 & 11.70 & 10.42 &  11 &       13 & $-3.98$ & $-4.50$ & $<33$  & $<-7.00$ & 2\\         
$0339352\!-\!352544$  & LP 944-20   & M9.0 & 10.75 &  9.52 &   5 &       26 & $-3.79$ & $-5.30$ & $74\pm 13$    & $-7.53$ & 7\\      
                      &             &      &       &       &     &          & $     $ &         & $2600\pm 200$ & $-5.99$ & 7\\
$0434152\!+\!225031$  &             & M9.0 & 13.74 & 11.87 & 140 &     \nod & $-2.53$ & \nod    & $<69$  & $<-5.92$ & 2\\         
$0436389\!+\!225812$  &             & M9.0 & 13.70 & 12.34 & 140 &     \nod & $-2.59$ & \nod    & $<57$  & $<-5.95$ & 2\\         
$0537259\!-\!023432$  &             & M9.0 & 18.22 & 17.00 & 352 &     \nod & $-3.56$ & \nod    & $<66$  & $<-4.11$ & 2\\         
$0810586\!+\!142039$  &             & M9.0 & 12.77 & 11.59 &  20 &       11 & $-3.39$ & \nod    & $<39$  & $<-6.52$ & 6\\           
$0853362\!-\!032932$  & GJ 3517     & M9.0 & 11.18 &  9.97 &   9 &     13.5 & $-3.49$ & $-3.93$ & $<81$  & $<-7.33$ & 5\\         
$1454280\!+\!160605$  & GJ 569Bb    & M9.0 & 11.65 & 10.43 &  10 &     \nod & $-4.04$ & \nod    & $<30$  & $<-7.08$ & 2\\         
$1627279\!+\!810507$  &             & M9.0 & 13.03 & 11.88 &  21 &  \nod    & $-3.45$ & \nod    & $<60$  & $<-6.23$ & 6\\           
$1707183\!+\!643933$  &             & M9.0 & 12.54 & 11.38 &  17 &  \nod    & $-3.44$ & \nod    & $<60$  & $<-6.43$ & 6\\        
$1707234\!-\!055824$  &             & M9.0 & 12.06 & 10.71 &  15 &     \nod & $-3.87$ & \nod    & $<48$  & $<-6.68$ & 2\\         
$0024246\!-\!015819$  & BRI B0021-0 & M9.5 & 11.99 & 10.54 &  12 &       33 & $-3.50$ & $-6.12$ & $83\pm 18$ & $-7.18$ & 5\\       
$0027420\!+\!050341$  & PC 0025+044 & M9.5 & 16.08 & 14.87 &  72 &       13 & $-3.62$ & $-3.39$ & $<75$  & $<-5.37$ & 5\\         
$0109217\!+\!294925$  &             & M9.5 & 12.91 & 11.68 &  19 &        7 & $-3.49$ & \nod    & $<54$  & $<-6.33$ & 6\\       
$0149089\!+\!295613$  &             & M9.5 & 13.45 & 11.98 &  17 &       12 & $-3.74$ & \nod    & $<140$ & $<-5.76$ & 6\\     
$0345431\!+\!254023$  &             & L0.0 & 13.92 & 12.67 &  27 &     \nod & $-3.56$ & \nod    & $<87$  & $<-6.22$ & 5\\         
$0746425\!+\!200032$  &             & L0.5 & 11.78 & 10.47 &  12 &       31 & $-3.93$ & $-5.29$ & $224\pm 15$     & $-6.38$ & 8\\      
                      &             &      &       &       &     &          &         &         & $15000\pm 100$ & $-4.55$ & 8\\
$1421314\!+\!182740$  &             & L0.0 & 13.23 & 11.94 &  20 &  \nod    & $-3.56$ & \nod    & $<42$  & $<-6.32$ & 6\\         
$0602304\!+\!391059$  & LSR 0602+39 & L1.0 & 12.30 & 10.86 &  11 &        9 & $-4.28$ & $-6.05$ & $<30$  & $<-6.78$ & 2\\       
$1300425\!+\!191235$  &             & L1.0 & 12.72 & 11.62 &  14 &       10 & $-4.12$ & $-5.71$ & $<87$  & $<-6.23$ & 2\\       
$0213288\!+\!444445$  &             & L1.5 & 13.49 & 12.21 &  19 &     \nod & $-4.24$ & \nod    & $<30$  & $<-6.32$ & 2\\       
$1807159\!+\!501531$  &             & L1.5 & 12.93 & 11.60 &  15 &       76 & $-4.24$ & $-5.26$ & $<39$  & $<-6.42$ & 2\\       
$2057540\!-\!025230$  &             & L1.5 & 13.12 & 11.72 &  16 &       62 & $-4.23$ & $-4.92$ & $<36$  & $<-6.37$ & 2\\       
$0109015\!-\!510049$  &             & L2.0 & 12.23 & 11.09 &  10 &     \nod & $-3.89$ & \nod    & $<111$ & $<-6.65$ & 3\\       
$0445538\!-\!304820$  &             & L2.0 & 13.41 & 11.98 &  17 &     \nod & $-4.33$ & \nod    & $<66$  & $<-5.99$ & 2\\       
$1305401\!-\!254110$  & Kelu-1      & L2.0 & 13.41 & 11.75 &  19 &       60 & $-3.57$ & $-5.69$ & $<27$  & $<-7.04$ & 4\\       
$0523382\!-\!140302$  &             & L2.5 & 13.08 & 11.64 &  13 &       21 & $-4.39$ & $-6.52$ & $<39$  & $<-6.61$ & 2\\       
                      &             &      &       &       &     &          &         &         & $231\pm 14$ & $-5.84$ & 2\\       
$0251149\!-\!035245$  &             & L3.0 & 13.06 & 11.66 &  12 &     \nod & $-4.42$ & \nod    & $<36$  & $<-6.12$ & 2\\       
$1721039\!+\!334416$  &             & L3.0 & 13.62 & 12.49 &  15 &     \nod & $-4.46$ & \nod    & $<48$  & $<-6.32$ & 2\\       
$2104149\!-\!103736$  &             & L3.0 & 13.84 & 12.37 &  17 &       27 & $-4.47$ & $-5.97$ & $<24$  & $<-6.26$ & 2\\       
$0036161\!+\!182110$  &             & L3.5 & 12.47 & 11.06 &   9 &       45 & $-4.51$ & $-6.26$ & $134\pm 16$ & $ -6.06$ & 5\\       
                      &             &      &       &       &     &          & $     $ &         & $720\pm 40$ & $ -5.33$ & 5\\       
$0045214\!+\!163444$  &             & L3.5 & 13.06 & 11.37 &  10 &     \nod & $-4.58$ & \nod    & $<39$  & $<-6.41$ & 2\\       
$1424390\!+\!091710$  &             & L4.0 & 15.69 & 14.17 &  32 &     \nod & $-4.04$ & \nod    & $<96$  & $<-5.56$ & 5\\       
$1705483\!-\!051646$  &             & L4.0 & 13.31 & 12.03 &  11 &       26 & $-4.65$ & $-7.12$ & $<45$  & $<-6.22$ & 2\\       
$0141032\!+\!180450$  &             & L4.5 & 13.88 & 12.49 &  13 &     \nod & $-4.76$ & \nod    & $<30$  & $<-6.15$ & 2\\       
$0652307\!+\!471034$  &             & L4.5 & 13.54 & 11.69 &  11 &     \nod & $-4.66$ & \nod    & $<33$  & $<-6.31$ & 2\\       
$2224438\!-\!015852$  &             & L4.5 & 14.07 & 12.02 &  11 &       32 & $-4.76$ & $-6.48$ & $<33$  & $<-6.19$ & 2\\       
$0004348\!-\!404405$  & GJ 1001BC   & L5.0 & 13.11 & 11.40 &  10 &       42 & $-4.67$ & $-7.42$ & $<45$  & $<-6.30$ & 2\\       
$0144353\!-\!071614$  &             & L5.0 & 14.19 & 12.27 &  13 &     \nod & $-4.73$ & \nod    & $<33$  & $<-6.08$ & 2\\       
$0205034\!+\!125142$  &             & L5.0 & 15.68 & 13.67 &  27 &     \nod & $-4.67$ & \nod    & $<48$  & $<-5.37$ & 2\\       
$0835425\!-\!081923$  &             & L5.0 & 13.17 & 11.14 &   9 &       23 & $-4.60$ & $-7.42$ & $<30$  & $<-6.58$ & 2\\       
$1228152\!-\!154734$  &             & L5.0 & 14.38 & 12.77 &  20 &       22 & $-4.19$ & \nod    & $<87$  & $<-5.84$ & 5\\       
$1507476\!-\!162738$  &             & L5.0 & 12.82 & 11.31 &   7 &       32 & $-4.23$ & $-8.18$ & $<57$  & $<-6.87$ & 5\\       
$1515008\!+\!484741$  &             & L6.0 & 14.06 & 12.56 &   9 &     \nod & $-5.11$ & \nod    & $<27$  & $<-6.12$ & 2\\       
$0439010\!-\!235308$  &             & L6.5 & 14.41 & 12.82 &  11 &     \nod & $-5.10$ & \nod    & $<42$  & $<-5.79$ & 2\\       
$0030300\!-\!145033$  &             & L7.0 & 16.28 & 14.48 &  27 &     \nod & $-5.01$ & \nod    & $<57$  & $<-4.96$ & 2\\       
$0205294\!-\!115929$  &             & L7.0 & 14.59 & 13.00 &  20 &       22 & $-4.65$ & \nod    & $<30$  & $<-5.87$ & 2\\       
$1728114\!+\!394859$  &             & L7.0 & 15.99 & 13.91 &  24 &     \nod & $-4.86$ & \nod    & $<54$  & $<-5.23$ & 2\\       
$0423485\!-\!041403$  &             & L7.5 & 14.46 & 12.93 &  15 &     \nod & $-4.83$ & \nod    & $<42$  & $<-5.77$ & 2\\       
$0825196\!+\!211552$  &             & L7.5 & 15.10 & 13.03 &  11 &       19 & $-5.21$ & $-8.18$ & $<45$  & $<-5.66$ & 2\\       
$2252107\!-\!173013$  &             & L7.5 & 14.31 & 12.90 &   8 &     \nod & $-5.29$ & \nod    & $<30$  & $<-5.98$ & 2\\       
$0929336\!+\!342952$  &             & L8.0 & 16.60 & 14.64 &  22 &     \nod & $-5.25$ & \nod    & $<42$  & $<-5.03$ & 2\\       
$1523226\!+\!301456$  &             & L8.0 & 16.06 & 14.35 &  19 &     \nod & $-5.27$ & \nod    & $<45$  & $<-5.12$ & 2\\       
$1632291\!+\!190441$  &             & L8.0 & 15.87 & 14.00 &  15 &       30 & $-5.31$ & \nod    & $<54$  & $<-5.17$ & 2\\       
$0151415\!+\!124430$  &             & T0.0 & 16.57 & 15.18 &  21 &     \nod & $-5.37$ & \nod    & $<51$  & $<-4.84$ & 2\\       
$2204105\!-\!564657$A &             & T1.0 & 12.29 & 11.35 &   4 &     \nod & $-5.03$ & \nod    & $<72$  & $<-6.58$ & 2\\       
$0207428\!+\!000056$  &             & T4.0 & 16.80 & 15.41 &  29 &     \nod & $-5.21$ & \nod    & $<39$  & $<-4.87$ & 2\\       
$0559191\!-\!140448$  &             & T4.5 & 13.80 & 13.58 &  10 &     \nod & $-4.53$ & \nod    & $<27$  & $<-6.60$ & 2\\       
$1534498\!-\!295227$  &             & T5.5 & 14.90 & 14.84 &  14 &     \nod & $-5.00$ & \nod    & $<63$  & $<-5.52$ & 2\\       
$1624144\!+\!002916$  &             & T6.0 & 15.49 & 15.52 &  11 &     \nod & $-5.16$ & \nod    & $<36$  & $<-5.78$ & 2\\       
$2204105\!-\!564657$B &             & T6.0 & 13.23 & 13.53 &   4 &     \nod & $-5.03$ & \nod    & $<72$  & $<-6.58$ & 2\\       
$1047539\!+\!212423$  &             & T6.5 & 15.82 & 16.41 &  11 &     \nod & $-5.35$ & \nod    & $<45$  & $<-6.26$ & 2\\       
$1346464\!-\!003150$  &             & T6.5 & 16.00 & 15.77 &  15 &     \nod & $-5.00$ & \nod    & $<105$ & $<-5.23$ & 5\\       
$0610351\!-\!215117$  & GJ 229B     & T7.0 & 14.20 & 14.30 &   6 &     \nod & $-5.21$ & \nod    & $<69$  & $<-6.01$ & 4\\       
$1217111\!-\!031113$  &             & T7.5 & 15.86 & 15.89 &  11 &     \nod & $-5.32$ & \nod    & $<111$ & $<-5.13$ & 2\\       
$0415195\!-\!093506$  &             & T8.0 & 15.70 & 15.43 &   6 &     \nod & $-5.73$ & \nod    & $<45$  & $<-5.68$ & 2       
\enddata 
\tablecomments{Properties of the M, L, and T dwarfs from the
literature.  The columns are (left to right): (i) 2MASS number; (ii)
other name; (iii) spectral type; (iv) $J$-band magnitude; (v) $K$-band
magnitude; (vi) distance from parallax or photometric estimate (from
SIMBAD, \citealt{fbc+09}, and \citealt{crl+03}); (vii) projected
rotation velocity (from \citealt{mb03,rkl+02,rb08,jrj+09,rb10});
(viii) bolometric luminosity; (ix) H$\alpha$ activity (from
\citealt{dfp+98,mb03,rb08,rb10}); (x) radio density flux; (xi) ratio
of radio to bolometric luminosity; and (xii) references for radio flux
density measurements: [1] \citet{wjk89}; [2] \citet{ber06}; [3]
\citet{bp05}; [4] \citet{kll99}; [5] \citet{ber02}; [6]
\citet{pol+07}; [7] \citet{bbb+01}; [8] \citet{brp+09}.}
\end{deluxetable}

\clearpage

\clearpage

\clearpage

\clearpage

\clearpage

\clearpage

\clearpage

\clearpage

\clearpage

\clearpage

\end{document}